\documentclass[useAMS,usenatbib]{mn2e}
\usepackage{graphicx}
\usepackage{subfigure}
\usepackage{hyperref}

\title[Periodic Eclipse Variations in V1432 Aql]{Periodic Eclipse Variations in Asynchronous Polar V1432 Aql: Evidence of a Shifting Threading Region}
\author[Littlefield et al.]{Colin Littlefield,$^{1, 2}$\thanks{E-mail:
clittlef@alumni.nd.edu.} Koji Mukai,$^{3, 4}$  Raymond Mumme,$^{1}$ Ryan Cain,$^{1}$\newauthor Katrina C. Magno,$^{1}$ Taylor Corpuz,$^{1}$ Davis Sandefur,$^{1}$ 
David Boyd,$^{5}$   Michael Cook,$^{6}$\newauthor Joseph Ulowetz,$^{7}$ Luis Martinez$^{8}$
\\ $^{1}$Department of Physics, University of Notre Dame, Notre Dame, IN 46556
\\ $^{2}$Department of Astronomy, Wesleyan University, Middletown, CT 06459
\\ $^{3}$CRESST and X-ray Astrophysics Laboratory, NASA/GCFC, Greenbelt MD 20771, USA
\\ $^{4}$Department of Physics, University of Maryland, Baltimore, MD 21250, USA
\\ $^{5}$CBA Oxford, 5 Silver Lane, West Challow, Wantage, OX12 9TX, United Kingdom
\\ $^{6}$CBA Ontario, Newcastle Observatory, 9 Laking Drive, Newcastle, Ontario, Canada L1B 1M5
\\ $^{7}$CBA Illinois, 855 Fair Lane, Northbrook, IL 60062
\\ $^{8}$Lenomiya Observatory, Casa Grande, Arizona}

\begin{document}

\date{Accepted ---. Received ---; in original form ---}

\pagerange{\pageref{firstpage}--\pageref{lastpage}} \pubyear{2015}

\maketitle

\label{firstpage}

\begin{abstract}
We report the results of a twenty-eight-month photometric campaign studying V1432 Aql, the only known eclipsing, asynchronous polar. Our data show that both the residual eclipse flux and eclipse O$-$C timings vary strongly as a function of the spin-orbit beat period. Relying upon a new model of the system, we show that cyclical changes in the location of the threading region along the ballistic trajectory of the accretion stream could produce both effects. This model predicts that the threading radius is variable, in contrast to previous studies which have assumed a constant threading radius. Additionally, we identify a very strong photometric maximum which is only visible for half of the beat cycle. The exact cause of this maximum is unclear, but we consider the possibility that it is the optical counterpart of the third accreting polecap proposed by \citet{rana}. Finally, the rate of change of the white dwarf's spin period is consistent with it being proportional to the difference between the spin and orbital periods, implying that the spin period is approaching the orbital period asymptotically.

\end{abstract}


\section{Introduction} \label{intro}
Cataclysmic variables (CVs) are interacting binary systems in which a low-mass star---usually a red dwarf---overfills its Roche lobe and transfers mass onto a white dwarf (WD).  \citet{warner} and \citet{hellier}  offer excellent overviews of these intriguing systems. In a subset of CVs known as polars, the exceptionally strong magnetic field ($\sim$ tens of MG) of the WD synchronizes the WD's spin period with the orbital period of the binary (see \citet{cropper} for a comprehensive review of polars specifically). The accretion stream from the secondary star follows a ballistic trajectory toward the WD until the magnetic pressure matches the stream's ram pressure. When this occurs, a threading region forms in which the accretion stream couples onto the WD's magnetic field lines, and the captured material is then channeled onto one or more accretion regions near the WD's magnetic poles. The impact of the stream creates a shock in which the plasma is heated to X-ray-emitting temperatures, so polars can be significantly brighter in X-ray wavelengths than ordinary non-magnetic CVs. In addition to X-rays, the accretion region produces polarized cyclotron emission in the optical and in the infrared, the detection of which is a defining characteristic of polars.

Eclipses of the WD have provided great insight into polars. Because a polar has no accretion disk, an eclipsing polar will generally exhibit a two-step eclipse: a very sharp eclipse of the compact ($\sim$ white dwarf radius) cyclotron-emitting region, followed by a much more gradual eclipse of the extended accretion stream (see, {\it e.g.}, \citet{harrop-allin} for an eclipse-mapping study of HU Aqr). When the accretion rate is high, the WD photosphere makes only a modest contribution to the overall optical flux, overshadowed by the two accretion-powered components mentioned above. 

Eclipsing polars also make it possible to determine the orientation of the magnetic axis with respect to the secondary.  In HU Aqr, the orientation of the dominant magnetic pole leads the line of centers of the binary by about 45$^{\circ}$ \citep{harrop-allin}, while in DP Leo, another eclipsing polar, the equilibrium orientation leads the line of centers by 7$^{\circ} \pm 3^{\circ}$ but with a long-term oscillation with an amplitude of $\sim25^{\circ}$ \citep{beuermann}.

In at least four polars,\footnote{In addition to the subject of this study (V1432 Aql), three other polars are incontrovertibly asynchronous: BY Cam, V1500 Cyg, and CD Ind. At the time of writing, there are at least two candidate systems: V4633 Sgr \citep{lipkin} and CP Pup \citep{bianchini}.} the WD's spin period differs from the orbital period by as much as several percent. In these asynchronous polars, the WD's magnetic field is gradually synchronizing the spin period with the orbital period on timescales of centuries. For example, \citet{ss91} detected a derivative in the WD spin period in V1500 Cyg and estimated that the system would approach resynchronization about 150 years after the publication of their study. 

Because the prototype asynchronous polar, V1500 Cyg, was almost certainly desynchronized during its 1975 nova eruption, the canonical view is that these systems are byproducts of nova eruptions which break the synchronous rotation by causing the primary to lose mass and to interact with the secondary \citep{ssl}. However, \citet{warner02} combined the fraction of asynchronous systems among all known polars with their estimated synchronization timescales and estimated an unexpectedly short nova recurrence time of a few thousand years for polars---far more rapid than the expected recurrence time of $\sim1 \times 10^5$ years. Every aspect of Warner's deduction ought to be explored, including the possibility of an additional channel for desynchronizing polars, selection effects that might alter the fraction of asynchronous polars, and methods of calculating the synchronization time scale. 

Interestingly, in each of the four confirmed asynchronous polars, the threading process is inefficient in comparison to fully synchronous systems. In synchronous systems, the accretion stream is fully captured not long after it leaves the L1 point, well before it can travel around the WD \citep[e.g.][]{schwope97}. In none of the asynchronous systems is this efficient threading seen. For example, Doppler tomography by \citet{schwope} of V1432 Aql showed an azimuthally extended accretion curtain, a finding which is possible only if the accretion stream can travel significantly around the WD. X-ray observations of V1432 Aql also indicate that the accretion stream travels most of the way around the WD before it is fully threaded onto the magnetic field lines \citep{mukai}. Likewise, in the other three systems, there is mounting evidence that the accretion flow can significantly extend around the WD. In CD Ind, the accretion stream appears to thread onto the same magnetic field line throughout the beat cycle, requiring that the stream be able to travel around the WD \citep{ramsay}. With regard to V1500 Cyg, \citet{ss91} argued that the smooth sinusoidal variation of the polarization curve was consistent with the infalling stream forming a thin accretion ring around the WD. More recently, \citet{litvinchova} detected evidence that this accretion ring is fragmented, periodically reducing the irradiation of the donor star by the hot WD. In the remaining system, BY Cam, Doppler tomograms show that the accretion curtain extends over $\sim180^\circ$ in azimuth around the WD, requiring a similar extent of the accretion stream \citep{schwarz}. Although a sample size of four is small, it is remarkable that in each of the confirmed asynchronous polars, the threading process is so inefficient that the accretion stream can travel much of the way around the WD. 

\section{V1432 Aql}

V1432 Aql (= RX J1940.1-1025) is the only known eclipsing, asynchronous polar and was identified as such by \citet{patterson} and \citet{friedrich}. There are two stable periodicities in optical photometry of V1432 Aql. The first (12116 seconds) is the orbital period, which is easily measured from the timings of the eclipses of the WD by the secondary. Initially, the nature of the eclipses was unclear; \citet{patterson} argued that the secondary was the occulting body, but \citet{watson} contended that a dense portion of the accretion stream was the culprit. Much of the confusion was attributable to the presence of residual emission lines and X-rays throughout the eclipses, as well as the variable eclipse depth. Since X-rays in polars originate on or just above the WD's surface, the apparent X-ray signal throughout the eclipse was inconsistent with occultations by the donor star. Additionally, there was considerable scatter in the eclipse timings, and the system's eclipse light curves did not show the rapid ingresses and egresses characteristic of synchronous polars \citep{watson}. However, \citet{mukai} resolved the dispute with high-quality X-ray observations which showed that the donor actually eclipses the WD and that the residual X-ray flux previously attributed to V1432 Aql was actually contamination from a nearby Seyfert galaxy.

The second periodicity ($\sim12150$ seconds) is the spin modulation of the WD. In optical photometry, this periodicity manifests itself in several ways. In particular, at $\phi_{sp} = 0.0$, the WD is occulted by material accreting onto one of the magnetic poles, producing a broad ``spin minimum'' \citep{friedrich}. Analyses of the spin minima have revealed several fascinating insights into V1432 Aql. For example, \citet{gs97} undertook an O$-$C study of the timing residuals of the spin minima and managed to detect a decrease in the WD spin period, indicating that the system is resynchronizing itself. They also measured a cyclical variation in the timings of the spin minima, caused by (1) a longitudinal offset between the magnetic pole and its corresponding accretion region on the WD's surface and (2) the accretion stream threading onto different magnetic field lines throughout the spin-orbit beat period  ($P^{-1}_{beat} = |P^{-1}_{orb}-P^{-1}_{sp}|$). Using these timings and a dipole accretion model, the authors managed to constrain the combined effect of the threading radius and the colatitude of the magnetic axis on the WD, but they could not constrain these parameters individually. \citet{staubert03} applied the methodology of \citet{gs97} to a larger dataset and refined the results of the earlier paper.

A critical concept which emerges from the literature is the beat period between the spin and orbital periods. The beat period is simply the amount of time that it takes for the WD (and its magnetic field) to rotate once as seen from the perspective of the donor star. As \citet{gs97} first demonstrated, the accretion stream will interact with different magnetic field lines as the system progresses through its beat period, a foundational principle which informs our analysis throughout this paper.

V1432 Aql is especially suitable for long-term study because its long-term brightness has remained constant not only in our own observations but also in data from the American Association of Variable Star Observers\footnote{www.aavso.org} dating back to 2002. Similarly, the Catalina Sky Survey \citep{drake} does not show any low states in the system since coverage of V1432 Aql began in 2005. While many polars alternate unpredictably between bright and faint states due to changes in the mass-transfer rate, V1432 Aql has not been observed to do so.

We supplement these previous studies by reporting the detection of stable periodicities in both the residual eclipse flux and the O$-$C timing residuals of the eclipses. These phenomena occur at the beat period, and we use a model to show that our observations are consistent with a threading radius whose position with respect to the WD varies throughout the beat cycle.

In response to this study's observational findings, one of us (DB) followed up by analyzing a different set of observations obtained by the Center for Backyard Astrophysics\footnote{http://cbastro.org/} over a much longer timespan. His group's analysis provides confirmation of the residual-flux and timing variations described in this paper while also reporting additional beat-cycle-related phenomena \citep{boyd}. 

\section{Observations}

\begin{figure}

	\includegraphics[width=0.45\textwidth]{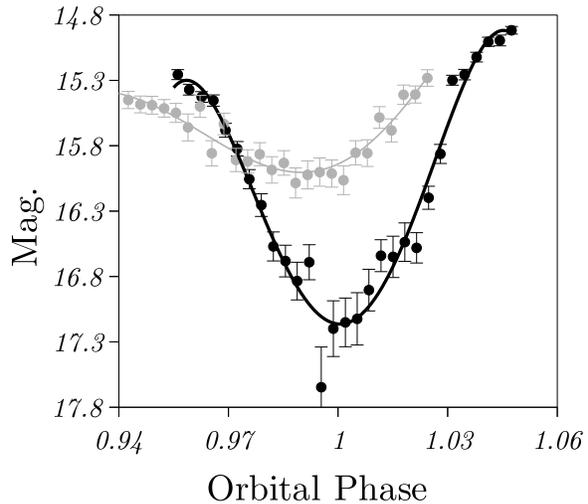}
	
\caption{Two representative eclipses of V1432 Aql. The data represented in black were obtained at $\phi_{beat} = 0.89$, and the data in gray at $\phi_{beat} = 0.54$. The solid lines are the best-fit polynomials for each dataset. The polynomials satisfactorily model the asymmetric eclipses while smoothing noisy, possibly spurious features in the light curves.}
\label{sample-eclipses}
\end{figure}

As part of a twenty-eight-month effort to study V1432 Aql's behavior at different beat phases, six of us (CL, RM, RC, KCM, TC, and DS) obtained unfiltered, time-resolved photometry using the University of Notre Dame's 28-cm Schmidt-Cassegrain telescope and SBIG ST-8XME CCD camera between July 2012 and July 2014. The exposure time was 30~seconds for each individual image, with an overhead time of 8~seconds per image. A total of 76 light curves, consisting of over 17,500 individual measurements, were obtained with this instrument. These observations constitute the bulk of our dataset, and their uniformity avoids the introduction of errors caused by combining unfiltered observations from different telescope-CCD combinations. Because of their homogeneity, we use these data for all three parts of our analysis: studying the eclipse O-C variations, measuring the mid-eclipse magnitude, and for constructing phase plots of the system at different beat phases.

We also obtained a number of light curves with other telescopes, but since these instruments have different spectral responses, we only used this supplemental data to explore eclipse O$-$C variations. CL obtained four unfiltered time series in July 2014 using the University of Notre Dame's 80-cm Sarah L. Krizmanich Telescope and two more with Wesleyan University's 60-cm Perkin Telescope in September 2014. The data obtained with the Krizmanich and Perkin Telescopes have much higher time resolution (exposure times between 5 and 7 seconds, each with a $\sim$3-second readout time, for a total cadence of 10 seconds or less), facilitating the study of the rapid variability during the eclipses. In addition, MC, JU, DB, and LM respectively used a 40-cm Schmidt-Cassegrain and QSI-516 CCD camera with a Johnson $V$ filter, a 23-cm Schmidt-Cassegrain and QSI-583ws CCD camera, a 25-cm Newtonian with an unfiltered SXV-H9 CCD camera, and a 28-cm Schmidt-Cassegrain equipped with an STT-1603 CCD camera. With the exception of LM, who used 45-second exposures, each of them used an exposure time of 60 seconds.

To compensate for light-travel delays caused by Earth's orbital motion, the timestamp for each observation was corrected to the BJD (TDB) standard \citep{eastman}.

With unfiltered photometry of a CV, it is possible to infer the approximate $V$-band magnitude of the CV by selecting a same-color comparison star and using its $V$ magnitude when calculating the magnitude of the CV. Since polars tend to be quite blue, we relied upon AAVSO field photometry to select two relatively comparison blue stars;\footnote{These stars are labeled 117 and 120 in AAVSO chart 13643GMF, and they have $B-V$ colors of 0.20 and 0.43, respectively, according to the APASS photometric survey \citep{APASS}.} we utilized these comparison stars for all photometry used in the analyses of mid-eclipse magnitude and the spin modulation at different beat phases.

One of the most obvious phenomena in the photometry is the highly variable magnitude of the system at mid-eclipse, which ranges from $V\sim$ 16.0 to $V\sim$ 17.5. Different eclipses also displayed strikingly different morphologies, and in Figure~\ref{sample-eclipses}, we plot two eclipses which are representative of this variation. Such behavior is plainly at odds with normal eclipsing polars, which almost invariably have very abrupt ingresses and egresses since most of the flux originates in a small---and thus rapidly eclipsed---area on the WD \citep[e.g.][]{harrop-allin}. V1432 Aql's gradual ingresses and egresses indicate that its flux originates in an extended region, and in this regard, its eclipses bear a superficial resemblance to those of CVs with accretion disks.

We measured both the time of minimum eclipse flux and the magnitude at mid-eclipse by fitting a fifth-order polynomial to each eclipse (see Table~\ref{eclipse_timings}). Figure~\ref{sample-eclipses} demonstrates the adequacy of the fit by plotting two eclipse light curves, each fitted with a fifth-order polynomial. Since the system's eclipses are frequently asymmetric, the time of minimum flux is not necessarily the midpoint between ingress and egress. Indeed, several eclipses were W-shaped, with two distinct minima. For these eclipses, we report the time of the deepest minimum. One particularly remarkable eclipse, observed on JD 2456843 and discussed in Section~\ref{application_of_model}, had two minima of equal depth, so we report both times.

Additionally, we detected a number of spin minima. Since previous studies of the spin minima \citep[e.g.][]{gs97} have measured the timing of each spin minimum by locating its vertical axis of symmetry, we fit a second-order polynomial to each spin minimum on the assumption that the minimum of this parabola will roughly approximate the vertical axis of symmetry. While a higher-order polynomial would do a better job of modeling the often-asymmetric spin minima, using the second-order polynomial increases the compatibility of our timings with those presented in other works.

We list in Table~\ref{spin_timings} the timings of all clearly-detected spin minima. A number of spin minima were ill-defined or had multiple mimima of comparable depth, and in those instances, we did not report a timing because it was impossible to objectively identify the middle of the spin minimum.

\begin{table}
	\centering
	\begin{minipage}{\textwidth}
	\caption{Observed Times of Minimum Eclipse Flux \label{eclipse_timings}}

	\begin{tabular}{cccccc}
	\hline
	BJD\footnote{$2456000+$} & $\phi_{beat}$ &$\phi_{sp}$& BJD& $\phi_{beat}$
	 & $\phi_{sp}$\\
	\hline

117.75353(52)&0.67&0.46&531.58818(26)&0.44&0.71\\
121.67928(60)&0.73&0.39&534.67318(47)&0.49&0.66\\
129.67477(51)&0.87&0.28&538.59720(41)&0.55&0.58\\
129.81416(50)&0.87&0.27&539.57897(30)&0.57&0.56\\
131.49702(40)&0.90&0.24&539.71935(49)&0.57&0.56\\
132.47891(47)&0.91&0.23&540.70141(35)&0.58&0.55\\
133.46253(73)&0.93&0.22&545.60953(54)&0.66&0.47\\
134.44280(55)&0.94&0.20&546.59071(53)&0.68&0.45\\
138.51071(58)&0.01&0.14&548.69482(44)&0.71&0.42\\
145.80112(28)&0.13&0.01&549.67629(38)&0.73&0.40\\
162.77047(40)&0.41&0.73&558.65164(39)&0.88&0.26\\
175.67045(38)&0.62&0.51&559.63391(37)&0.89&0.24\\
180.57845(40)&0.70&0.43&560.61564(49)&0.91&0.23\\
180.71866(44)&0.70&0.43&562.58006(36)&0.94&0.21\\
181.70019(43)&0.72&0.41&565.66515(61)&0.99&0.16\\
182.68111(44)&0.74&0.39&566.64823(74)&0.01&0.15\\
194.60329(44)&0.93&0.21&567.62862(61)&0.02&0.12\\
428.79371(28)&0.76&0.37&573.65832(71)&0.12&0.02\\
431.87824(79)&0.81&0.31&574.63842(48)&0.14&1.00\\
447.86614(30)&0.07&0.06&575.61915(43)&0.15&0.97\\
451.79366(39)&0.14&0.00&576.60209(40)&0.17&0.97\\
460.76848(41)&0.28&0.85&577.58372(37)&0.18&0.95\\
462.73257(35)&0.32&0.83&579.54661(33)&0.22&0.92\\
463.71538(66)&0.33&0.82&580.52958(50)&0.23&0.91\\
477.73569(51)&0.56&0.57&593.57155(31)&0.44&0.70\\
484.74744(37)&0.68&0.45&594.55383(40)&0.46&0.69\\
484.74771(71)&0.68&0.46&600.58074(45)&0.56&0.57\\
484.88740(65)&0.68&0.45&787.79453(54)&0.58&0.54\\
485.72961(41)&0.69&0.44&799.85494(46)&0.78&0.35\\
485.72974(50)&0.69&0.44&801.81835(58)&0.81&0.32\\
486.71075(55)&0.71&0.42&813.73962(81)&0.00&0.14\\
486.85131(39)&0.71&0.42&814.72096(43)&0.02&0.12\\
486.85137(40)&0.71&0.42&815.70217(53)&0.03&0.10\\
487.69245(51)&0.72&0.41&815.8433(12)&0.04&0.10\\
487.69267(51)&0.72&0.41&822.85472(77)&0.15&0.99\\
488.81394(44)&0.74&0.39&842.76944(26)&0.47&0.68\\
490.77791(51)&0.77&0.36&842.76957(49)&0.47&0.68\\
503.67970(72)&0.98&0.15&843.74858(55)&0.48&0.65\\
506.62498(99)&0.03&0.10&843.75227(55)&0.48&0.67\\
506.76450(79)&0.03&0.10&843.7524(12)&0.48&0.67\\
508.72870(47)&0.07&0.07&847.67489(53)&0.55&0.58\\
510.69232(55)&0.10&0.04&847.8138(15)&0.55&0.57\\
515.60006(51)&0.18&0.96&848.65653(31)&0.56&0.56\\
528.64310(38)&0.39&0.76&849.77851(33)&0.58&0.55\\
529.62302(21)&0.40&0.73&903.77121(45)&0.45&0.70\\
529.76522(39)&0.41&0.74&904.61311(18)&0.46&0.69\\
530.60621(49)&0.42&0.72&905.59470(33)&0.48&0.67\\

	\hline
	\end{tabular}

	\end{minipage}
\end{table}

\section{Analysis}

\subsection{Orbital, Spin and Beat Ephemerides}\label{ephem}

	\begin{figure}

	\includegraphics[width=0.45\textwidth]{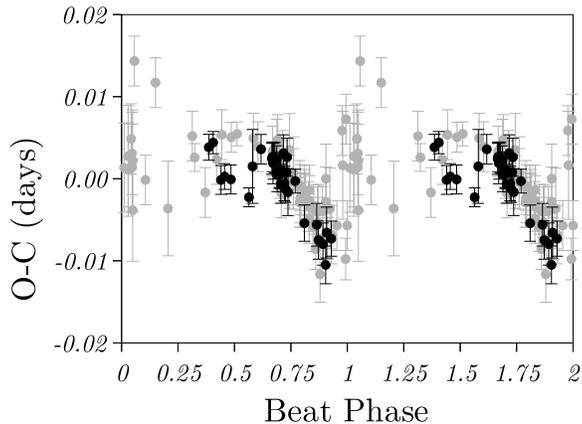}
	\caption{O$-$C timing residuals for the spin minima as a function of $\phi_{beat}$. The black dataset represents the new timings which we report in Table~\ref{spin_timings}, while the gray datapoints are from previously published studies as described in the text. The data are repeated for clarity. Our lack of timings from $0.0 < \phi_{beat} < 0.5$ is a consequence of the weakness of the spin minima during this half of the beat cycle.}
	\label{minima-timings}
	
	\end{figure}

We used $\chi^{2}$ minimization to determine the best-fit ephemerides for the spin and orbital periods using our data in conjunction with the published optical eclipse and spin-minima timings in \citet{patterson}, \citet{gs97}, \citet{staubert03}, and \citet{mukai}. Some of the timings from these studies lacked uncertainties; for those observations, we adopted the average uncertainty of all measurements which did have error estimates. Furthermore, both \citet{abb06} and \citet{b12} have made their photometry of V1432 Aql available electronically, and while their time resolution was too low for inclusion in our eclipse analysis, it was adequate for measuring the spin minima. In the interest of uniformity of analysis, we measured the spin minima in the \citet{abb06} and \citet{b12} datasets ourselves instead of using their published timings.\footnote{The original preprint of this paper used the timings reported in \citet{b12} without reanalyzing their photometry. Using our timing measurements of their spin minima resulted in a significantly lower values of values of $\chi^{2}_{red}$ for our spin ephemerides in Section~\ref{spin_ephemerides}.}

\subsubsection{Orbital Ephemeris}

The best-fit linear eclipse ephemeris is \begin{equation} T_{ecl}[HJD] = T_{0, ecl} + P_{orb}E_{ecl},\end{equation} with $T_{0, ecl} = 2454289.51352 \pm 0.00004$ and  $P_{orb} = 0.1402347644 \pm 0.0000000018$ d. Even though our timestamps use the BJD standard, we report our epochs using the slightly less accurate HJD standard because the previously published data use HJD. We find no evidence of a period derivative in the orbital ephemeris, but both \citet{b12} and \citet{boyd} have reported quadratic orbital ephemerides. The latter paper had a larger dataset than the one used in this study, so our non-detection of an orbital period derivative does not necessarily contradict those claims.

\subsubsection{Spin Ephemeris}\label{spin_ephemerides}

The spin ephemeris of \citet{b12} fits our data very well, and we offer only a modestly refined cubic spin ephemeris of \begin{equation} T_{min, sp}[HJD] = T_{0, sp} + P_{sp,0}E_{sp} + \frac{\dot{P}}{2}E^{2}_{sp} + \frac{\ddot{P}}{6}E^{3}_{sp},\end{equation} where $T_{min}$ is the midpoint of the spin minimum, $T_{0, sp} =  2449638.3278 (\pm 0.001), P_{sp,0} = 0.14062835 (\pm 0.00000022)$ d, $\dot{P}/2 = -8.10 (\pm 0.10) \times 10^{-10}$ d cycle$^{-2}$, and $\ddot{P}/6 =  -8.5 (\pm 1.4) \times 10^{-16}$ d cycle$^{-3}$. The uncertainties on these parameters were determined by bootstrapping the data. We do not have enough observations to meaningfully search for higher-order period derivatives like those reported by \citet{boyd}, but these values are within the error bounds of those reported by \citet{b12}. 

While a polynomial fit accurately models the existing data, $P_{sp}$ will likely approach $P_{orb}$ asymptotically over the synchronization timescale (P. Garnavich, private communication). If this is correct, then $\dot{P}$ is probably proportional to the difference between $P_{sp}$ and $P_{orb}$ so that \begin{equation}\dot{P} \equiv \frac{dP_{sp}}{dE_{sp}} = k(P_{sp} - P_{orb}).\label{pdot_exp}\end{equation} Integrating the solution to this differential equation yields an ephemeris of \begin{equation} T_{min, sp} = \frac{P_{sp, 0} - P_{orb}}{k}(e^{kE_{sp}} - 1) + P_{orb}E_{sp} + T_{0, sp}, \end{equation} where $P_{orb}$ is the measured value and the three free parameters are $k = -4.205 (\pm0.008) \times 10^{-6}$ cycles$^{-1}$, $P_{sp, 0} = 0.14062863 (\pm0.00000008)$ d, and $T_{0} = 2449638.3277 (\pm0.0010)$. 

Although $\chi^{2}_{red} = 2.9$ for both the cubic ephemeris and the exponential ephemeris, both of these ephemerides  neglect the cyclical shifts in the location of the accretion spot first reported by \citet{gs97}. To illustrate the effect of these variations on the quality of our fit, Figure~\ref{minima-timings} plots the residuals from the cubic ephemeris as a function of beat phase. Because this particular variation is not an actual change in the spin period, we did not attempt to incorporate it into our ephemerides. Unless a spin ephemeris were to take into account these variations and their $\sim$1000-second peak-to-peak amplitude, it would be difficult to achieve a significantly lower $\chi^{2}_{red}$.

With this caveat in mind, the comparable values of $\chi^{2}_{red}$ for each ephemeris lead us to conclude that they model the data equally well as could be expected. Though we use the cubic ephemeris for the sake of simplicity when calculating the beat phase, the exponential spin ephemeris is at least grounded in a physical theory of the resynchronization process. Moreover, in principle, the only parameter which should need to be updated in the future is the constant $k$. By contrast, a polynomial ephemeris could require an ungainly number of terms in order to attain a satisfactory fit.

\subsubsection{Beat Ephemeris}

Because there are several non-trivial steps in calculating the system's beat phase, the beat ephemeris is too unwieldy to list here. Nevertheless, to facilitate future studies, we have written a Python script which calculates the system's beat phase at a user-specified Heliocentric Julian Date using the procedure outlined in Appendix~\ref{beatphase}. Additionally, it calculates future dates at which the system will reach a user-specified beat phase. This script is available for download as supplemental online material and may also be obtained via e-mail from CL.

\subsubsection{Synchronization Timescale}

As defined by \citet{ss91}, a first-order approximation of an asynchronous polar's synchronization timescale is given by \begin{equation} \tau_{s} = \frac{P_{orb} - P_{sp}}{\dot{P}}.\label{timescale-formula} \end{equation} If one assumes rather unrealistically that $\dot{P}$ will remain constant until resynchronization, this formula provides a very rough estimate of when resynchronization will occur. If Equation~\ref{pdot_exp} is substituted for $\dot{P}$ in Equation~\ref{timescale-formula}, this equation simplifies to $\tau_{s} = -k^{-1}$. Since $k$ is essentially a decay rate, this formula yields the amount of time necessary for the initial value (in this context, the asynchronism at $T_0$, given by $P_{spin,0} - P_{orb}$) to be reduced by a factor of $e^{-1}$. Because $-k^{-1} = 237700$ spin cycles, $\tau_{s} = 71.5\pm0.4$ years with respect to August 2014, so in the year $\sim$2086, the predicted spin period would be $\sim$12128.8 seconds, fully 12.5 seconds longer than $P_{orb}$. While this estimate of $\tau_{s}$ is obviously not an estimate of when resynchronization will actually occur, it is slightly less than the values in \citet{gs97} and \citet{abb06} and considerably less than \citet{staubert03}.

It is unclear how long an exponential spin ephemeris might remain valid, but if ours were to hold true indefinitely, it predicts that $P_{sp}$ will approach $P_{orb}$ to within one second in the year $\sim2320$ and to within 0.1 seconds in $\sim2750$. These are not synchronization timescales as defined by \citet{ss91}, but in the case of an exponential ephemeris, they provide a more realistic manner of extrapolating when the system might approach resynchronization. The inferred $\sim$300-year timespan necessary just to attain $P_{sp} - P_{orb} < 1$ seconds is longer than the $\sim$100-year timescales in \citet{gs97} and \citet{abb06}, but it is within the error bounds of the $\sim$200-year synchronization timescale announced in \citet{staubert03}. An important disclaimer with these synchronization timescales is that the orbital period may be decreasing, as claimed by both \citet{b12} and \citet{boyd}. Since V1432 Aql's WD is spinning up, a decreasing orbital period would presumably lengthen the resynchronization timescale.

If asynchronous polars do resynchronize asymptotically, it would suggest that a number of supposedly synchronous polars are very slightly asynchronous, with beat periods of months, years, or even decades. Unless they were closely observed for extended periods of time, these polars might be misclassified as being synchronous, so the true fraction of polars which are asynchronous might actually be higher than is currently believed. If correct, this result would be relevant in any examination of the problem of the unreasonably short nova-recurrence time in polars \citep{warner02}. On one hand, a greater proportion of polars which are asynchronous would imply an even faster recurrence time, but on the other hand, an asymptotic approach to synchronism would also prolong the resynchronization process---and thus, the recurrence time. We leave it to a future work to more fully explicate these matters, but clearly, it will be important to independently confirm our exponential ephemeris, to resolve the possibility of an orbital-period derivative in V1432 Aql, and to determine if the other asynchronous systems also show evidence of asymptotic resynchronization.

\subsection{Variations in Eclipse O$-$C} \label{O-C}

\begin{figure*}

	\begin{subfigure}{
	\includegraphics[width=0.5\textwidth]{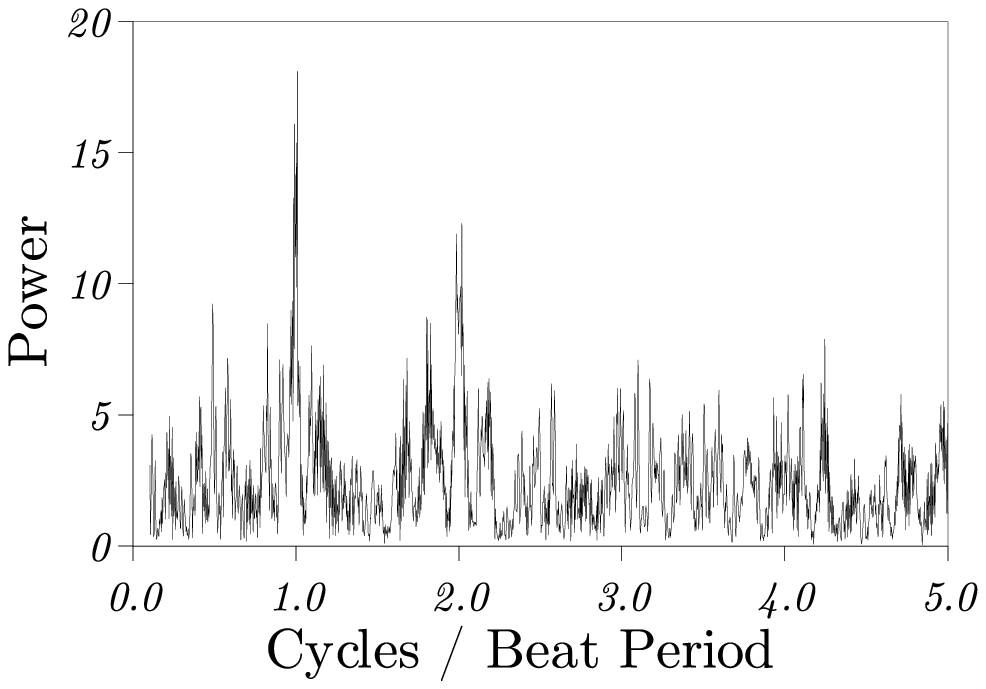}
	\includegraphics[width=0.5\textwidth]{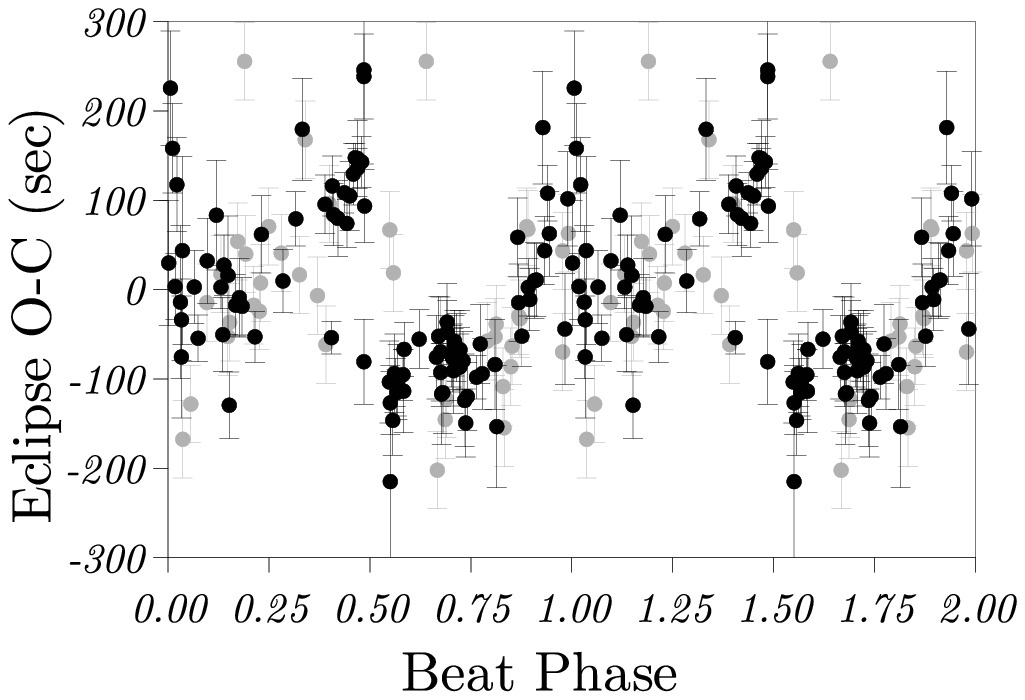}}
	\end{subfigure}
	
\caption{From left to right: the power spectrum of the timing residuals of the combined dataset described in section~\ref{O-C}, and the waveform of the combined dataset when phased at the beat period. Black data points represent our data as listed in Table~\ref{eclipse_timings}, while gray data points indicate previously published data as described in Section~\ref{O-C}.}
\label{timing}
\end{figure*}

\subsubsection{Periodicity}

In a conference abstract, \citet{gs99} first reported the discovery of a 200-second O$-$C shift in V1432 Aql's eclipse timings. We followed up on this periodicity by performing an O$-$C analysis on all eclipse timings listed in Table~\ref{eclipse_timings}. We calculated both the O$-$C timing residual and the beat cycle count ($C_{beat}$; see Appendix~\ref{beatphase}) for each eclipse and then used the analysis-of-variance (ANOVA) technique \citep{anova} to generate several periodograms, with $C_{beat}$ serving as the abscissa.

The first periodogram used all of the eclipse timings reported in Table~\ref{eclipse_timings}, and it showed a moderately strong signal at 1.00$\pm$0.02 cycles per beat period, with the folded eclipse timings exhibiting a sawtooth waveform. We then recalculated the power spectrum after adding previously published optical eclipse timings by \citet{patterson} and \citet{watson} to the dataset. 

\begin{table}
	\centering

	\begin{minipage}{\textwidth}
	\caption{Observed Times of Spin Minima\label{spin_timings}}
	\begin{tabular}{ccc|ccc}
	\hline
	BJD\footnote{2456000+} & $\phi_{beat}$ & $\phi_{orb}$ & BJD & $\phi_{beat}$ & $\phi_{orb}$\\
	\hline

117.8317(18)&0.64&0.55&486.7913(24)&0.69&0.57\\
119.6574(22)&0.67&0.57&486.7925(18)&0.69&0.58\\
121.6249(21)&0.70&0.60&487.7752(21)&0.70&0.58\\
129.7710(26)&0.84&0.69&488.7581(29)&0.72&0.59\\
131.4553(26)&0.87&0.70&490.7272(15)&0.75&0.63\\
132.4406(31)&0.88&0.73&528.6812(16)&0.37&0.28\\
162.6717(14)&0.38&0.3&534.7212(18)&0.47&0.35\\
175.6023(18)&0.59&0.51&539.6384(12)&0.55&0.41\\
180.6599(21)&0.68&0.57&540.6261(45)&0.57&0.46\\
181.6459(22)&0.69&0.61&546.6710(19)&0.66&0.56\\
182.6293(23)&0.71&0.62&549.6188(16)&0.71&0.58\\
194.5668(21)&0.90&0.74&558.6081(23)&0.86&0.69\\
431.8292(22)&0.79&0.64&560.5729(23)&0.89&0.70\\
484.6855(25)&0.65&0.55&593.6136(18)&0.43&0.31\\
484.8263(22)&0.66&0.55&594.5979(16)&0.44&0.32\\
485.6698(17)&0.67&0.57&607.5310(19)&0.65&0.55\\
485.8085(21)&0.67&0.56&&&\\

	\hline
	\end{tabular}

\end{minipage}

\end{table} 

The combined dataset consists of 133 measurements spanning a total of 138 beat cycles.  The strongest signal is at the beat period (1.001$\pm$0.002 cycles per beat period), and its waveform consists of an abrupt 240-second shift in the timing variations near $\phi_{beat}\sim0.5$, which is when the residual eclipse flux is strongest (see Section~\ref{flux-periodicity}). Both the periodogram and waveform are shown in Figure~\ref{timing}. Between $\sim0.5 < \phi_{beat} < \sim 0.85$, the eclipses occur $\sim$120 seconds early, but after $\phi_{beat} \sim 0.85$, the eclipses begin occurring later, and by $\phi_{beat} \sim 1.0$, the eclipses are occurring $\sim$120 seconds late. Although the 240-second O$-$C jump at $\phi_{beat}\sim0.5$ is the most obvious feature in the O$-$C plot, there is a 120-second jump towards earlier eclipses at $\phi_{beat}\sim0.0$. Considering the gradual eclipse ingresses and egresses, the WD must be surrounded by an extended emission region, so these eclipse timings track the centroid of emission rather than the actual position of the WD.

\subsubsection{Description of Model} \label{description_of_model}

Given the asynchronous nature of the system and the ability of the stream to travel most of the way around the WD \citep{mukai}, we hypothesize that cyclical changes in the location of the threading region are responsible for the O$-$C variation. In an asynchronous system, the position of the threading region can vary because the WD rotates with respect to the accretion stream, causing the amount of magnetic pressure at a given point along the stream to vary during the beat period. Threading occurs when the magnetic pressure ($\propto r^{-6}$) balances the stream's ram pressure  ($\propto v^{2}$). For a magnetic dipole, the magnetic flux density $B$ has a radial dependence of $\propto r^{-3}$, but with an additional dependence on the magnetic latitude; the magnetic pressure will be even greater by a factor of 4 near a magnetic pole as opposed to the magnetic equator. An additional consideration is that the stream's diameter is large enough that the magnetic pressure varies appreciably across the stream's cross section \citep{mukai88}.

KM modeled this scenario using a program which predicts times of eclipse ingresses and egresses of a point given its $x, y$, and $z$ coordinates within the corotating frame of the binary. The physical parameters used in the program are $P_{orb} = 3.365664$ h (measured), $M_{WD}$ = 0.88M$_{\odot}$, $M_{donor}$ = 0.31M$_{\odot}$, $R_{donor}$ = $2.47 \times 10^{10}$ cm, $i = 76.8^{\circ}$, and binary separation $a = 8.4 \times 10^{10}$ cm \citep{mukai}. The code treats the donor star as a sphere for simplicity, but since we do not attempt to comprehensively model the system in this paper, the errors introduced by this approximation should be minimal. For instance, as a result of this approximation, we had to decrease $i$ by 0.9$^{\circ}$ compared to the value from \citet{mukai} in order to reproduce the observed eclipse length.

\begin{figure}

	\includegraphics[width=0.5\textwidth]{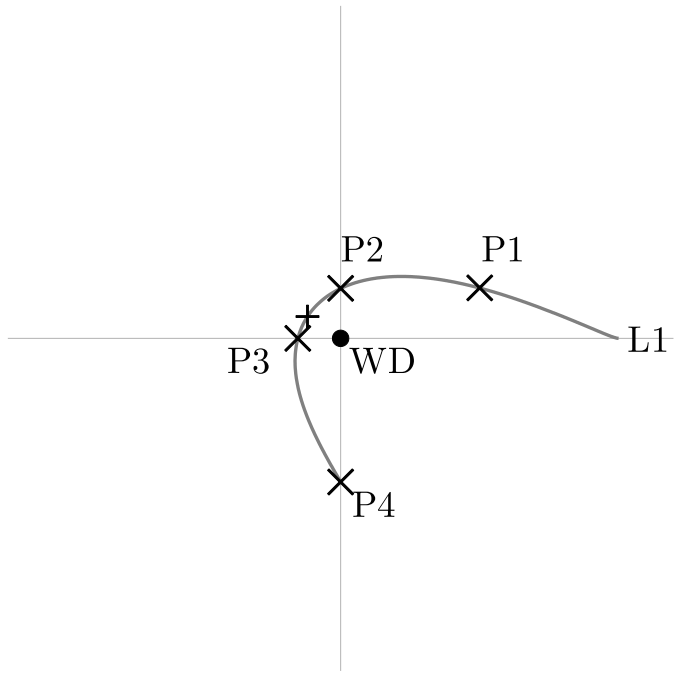}
	\caption{A schematic diagram of the system as used in our model, viewed from above the binary rest frame. The WD is rest at the origin, and the black curved line is the accretion stream trajectory, which originates at the L1 point near the right edge of the diagram. P1, P2, P3, and P4 are illustrative threading regions, and the cross indicates the location of the stream's closest approach to the WD. Since $P_{sp} > P_{orb}$, the WD rotates clockwise in this figure.}
	\label{model}
	
	\end{figure}

We first calculated the ballistic trajectory of the accretion stream and arbitrarily selected four candidate threading regions along the stream (P1, P2, P3, and P4) under the assumption that the stream will follow its ballistic trajectory until captured by the magnetic field \citep{mukai88}. The eclipse-prediction program then returned the phases of ingress and egress for each of the four points given their $x$ and $y$ coordinates within the corotating frame of the binary. We selected these four points arbitrarily in order to demonstrate the effects that a changing threading region would have on eclipse O$-$C timings; we do not claim that threading necessarily occurs at these positions or that this process is confined to a discrete point in the $x,y$ plane. Figure~\ref{model} shows a schematic diagram of this model.

\begin{figure*}

	\includegraphics[width=1\textwidth]{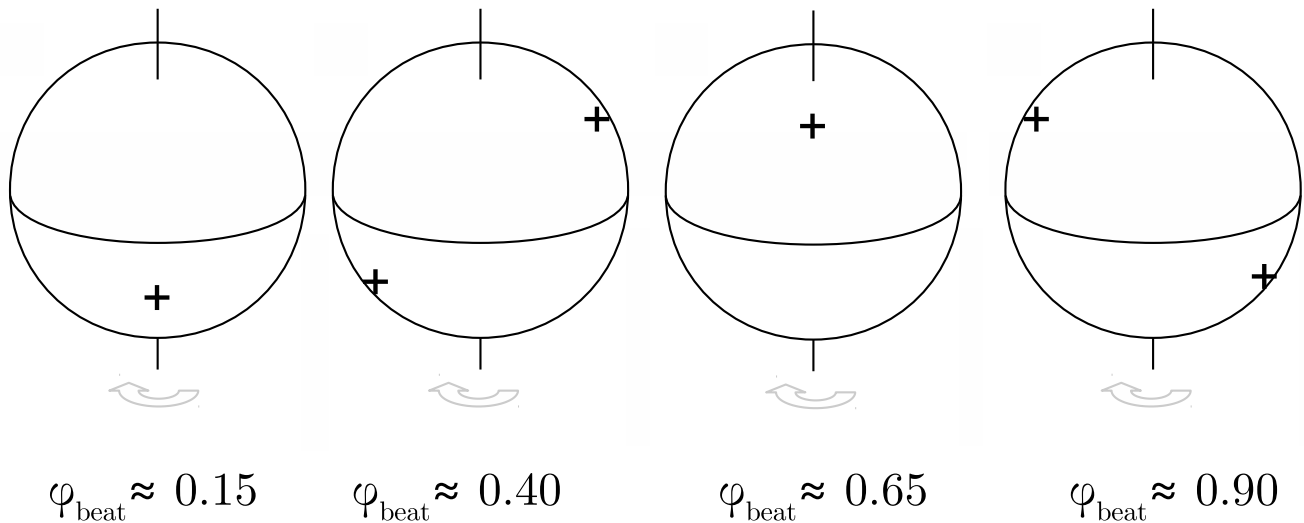}
	
\caption{A sketch indicating the general positions of the accretion spots at different beat phases as seen from the donor star. The black crosses represent accretion spots visible from the donor, and the vertical line is the WD's spin axis. Section~\ref{orientation} explains how we inferred the positions of the two magnetic poles.}
\label{diagram}
\end{figure*}

Once threading occurs, the captured material will follow the WD's magnetic field lines until it accretes onto the WD. To simulate the magnetically channeled portion of the stream, we assumed that captured material travels in a straight line in the $x,y$ plane from the threading region to the WD while curving in the $z$ direction, where $z$ is the elevation above or below the $x,y$ plane. This is another simplification since the magnetic portion of the stream might be curved in the $x,y$ plane, but presumably, this approximation is reasonable. Since the magnetic field lines will lift the captured material out of the orbital plane, we calculated the $x,y$ coordinates of the midpoint between each threading region and the WD and computed its ingress and egress phases at several different values of $z$. We reiterate that this is not a comprehensive model, but as we explain shortly, it is sufficiently robust to offer an explanation for the observed O$-$C variations.

\subsubsection{Orientation of the Poles} \label{orientation}

Before this model is applied to the observations, it is helpful to determine the orientations of the poles at different points in the beat cycle. We assume that there are two magnetic poles which are roughly opposite each other on the WD \citep{mukai}. Since $i \neq 90^{\circ}$, one hemisphere of the WD is preferentially tilted toward Earth, and we refer to the magnetic pole in that hemisphere as the upper pole. The lower pole is the magnetic pole in the hemisphere which is less favorably viewed from Earth. In isolation, our observations do not unambiguously distinguish between these two poles, but since the midpoint of the spin minimum ({\it i.e.},  $\phi_{sp} = 0.0$) corresponds with the transit of the accretion region across the meridian of the WD \citep[e.g.][]{staubert03}, we can estimate when the poles face the donor star. When $\phi_{beat} \sim 0.15$, the spin minimum coincides with the orbital eclipse, so one of the poles is approximately oriented towards the secondary at that beat phase. At $\phi_{beat} \sim 0.65$, the spin minimum occurs at an orbital phase of $\sim$0.5, indicating that this pole is roughly facing the P3 region at that beat phase. But the question remains: Is this the upper pole, or the lower one?

\citet{mukai} relied upon X-ray observations of eclipse ingresses and egresses to differentiate between the upper and lower poles (see their Figure~15 and the accompanying text). While the accretion spots have not been identified in optical photometry, they are the system's dominant X-ray sources, so they produce steep, rapid X-ray eclipses \citep{mukai}. The authors took advantage of the fact that since $P_{sp} > P_{orb}$,  the accretion spots will increasingly lag  behind the orbital motion of the donor star with each subsequent orbit. Consequently, the orientation of the accretion regions with respect to the donor star will continuously change across the beat cycle. When viewed throughout the beat period at the phase of eclipse, the accretion spots appear to slowly move across the face of the WD, thereby causing detectable changes in the times of X-ray ingress and egress. 

Critically, at some point during the beat cycle, each accretion region will have rotated out of view at the phase of eclipse, resulting in a jump in either the ingress or egress timings, depending on which pole has disappeared. The \citet{mukai} model predicts that when the upper pole is aimed in the general direction of P4, the X-ray egresses will undergo a shift to later phases as the upper polecap rotates behind the left limb of the WD as seen at egress (see their Figure~15). Likewise, the disappearance of the lower pole behind the left limb at the phase of ingress results in a shift toward later phases in the ingress timings. Based on data in Table~5 of \citet{mukai}, the egress jump occurs near $\phi_{beat}\sim0.9$, so at that beat phase, the upper pole should be pointed toward the P3-P4 region. The egress jump is more distinct than the ingress jump, so we base our identification of the poles on the egress jump only.

Our identification of the upper and lower poles is an inference and should not be viewed as a definitive claim. For our method to be reliable, it would be necessary for the accretion geometry to repeat itself almost perfectly in both 1998 (when \citet{mukai} observed) and the 28-month span from 2012-2014 when we observed V1432 Aql. Even though the accretion geometry does seem to repeat itself on a timescale of two decades (see, {\it e.g.}, Section~\ref{spin}), this may not always be the case, as is evidenced by an apparent discontinuity in the timings of the of the spin minima in 2002 \citep{boyd}. If the accretion rate during our observations was different than it was in 1998, there would be changes in the location and size of the X-ray-emitting accretion regions \citep{mukai88}. Moreover, \citet{mukai} cautioned that their model was a simplification because the accretion geometry was poorly constrained. For example, they noted that their model did not account for the offset between the accretion region and the corresponding magnetic pole.

If the upper pole is aimed towards P3-P4 near $\phi_{beat}\sim0.9$, then the upper pole would face the donor at $\phi_{beat} \sim 0.65$ since the WD appears to rotate clockwise as seen from the donor. Thus, the lower pole is likely pointed in the general direction of the donor star near $\phi_{beat} \sim 0.15$. We provide a sketch of the system in Figure~\ref{diagram} which shows the inferred positions of the polecaps throughout the beat cycle.

\subsubsection{Application of Model}\label{application_of_model}

\begin{figure}

	\includegraphics[width=0.45\textwidth]{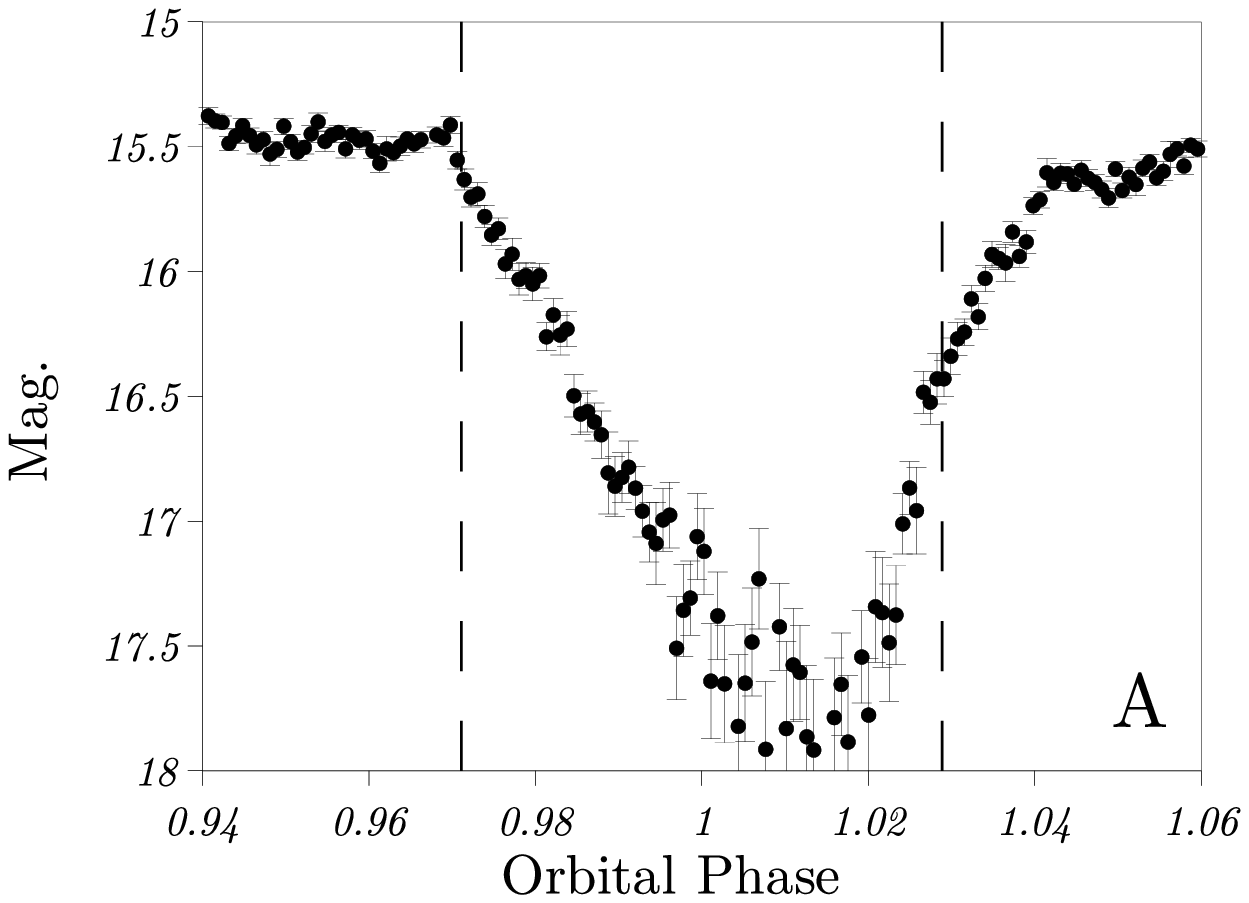}
	\includegraphics[width=0.45\textwidth]{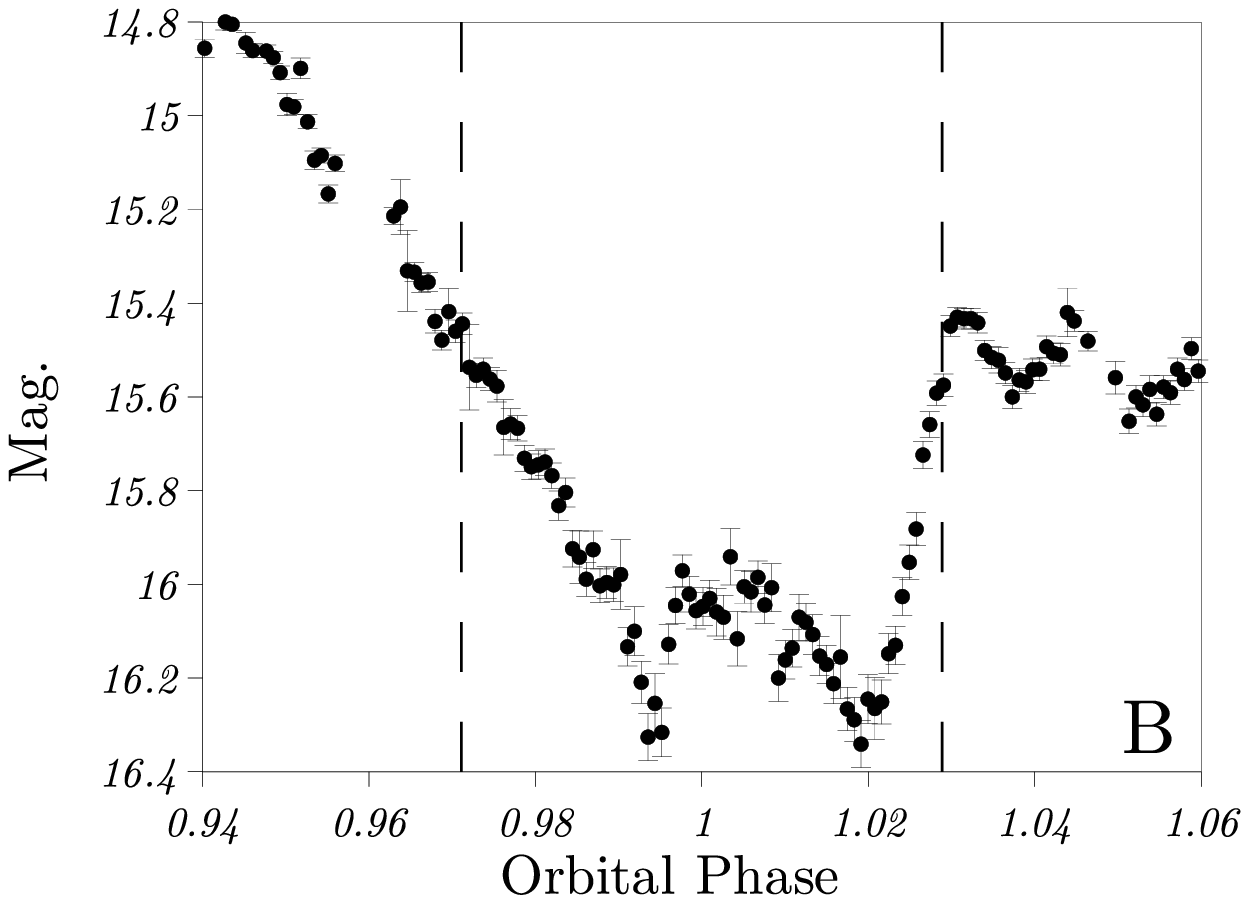}
	
\caption{Two eclipses observed on consecutive nights with the 80-cm Krizmanich Telescope. Note the different vertical scale for the two panels. The vertical dashed lines indicate the expected phases of the WD's ingress and egress. On the first night (Panel A), the eclipse is very deep and begins with the WD's disappearance, but on the second night (Panel B), the eclipse starts before the occultation of the WD. These light curves are consistent with the appearance of a new threading region near P3-P4 in our model, indicating that this process requires less than 24 hours to take place.}
\label{shift}
\end{figure}

Even though the four P points were arbitrarily selected, the results of the eclipse-timing program provide testable predictions concerning the O$-$C variations. In our model, the emission from the accretion curtain and the threading region result in a moving centroid which is responsible for an O$-$C shift with a half-amplitude of about $\pm$120 seconds (see Fig.~\ref{O-C}). When the centroid of emission is in the $+y$ region in Figure~\ref{model}, the O$-$C would be positive, and if it were in the $-y$ half of the plot, the O$-$C would be negative. According to calculations using the model, eclipses of point sources at P1, P2, P3, and P4 would result in O$-$C values of 289 seconds, 204 seconds, 0 seconds, and $-$533 seconds, respectively. As for the midpoints between each of those four points and the WD, the O$-$C values would be 122 seconds, 103 seconds, 0 seconds, and $-$289 seconds for the P1, P2, P3, and P4 midpoints, respectively. The O$-$C values for the midpoints have a negligible dependence on the height above the orbital plane (provided that the secondary can still eclipse that point). Since the actual O$-$C variation does not exceed $\pm$120 seconds, it is clear that the actual O$-$C timings are inconsistent with a centroid near P1, P2, and P4. However, centroids near the midpoints for P1, P2, and P3 would be consistent with the observed O$-$C timings.

It makes sense that the centroid of the emission region would have a less dramatic O$-$C value than the candidate threading points. Because we expect that the magnetically-channeled part of the stream travels from the threading region to the WD, the light from this accretion curtain would shift the projected centroid of emission towards the WD. In addition, since the threading region likely subtends a wide azimuthal range, the ability of the projected centroid to deviate dramatically from the WD's position would be limited. With these considerations in mind, the consistency of the theoretical O$-$C values for the P1, P2, and P3 midpoints with the observed O$-$C variations indicates that our model offers a plausible explanation of the O$-$C timings.

The sudden jump to early eclipses near $\phi_{beat} \sim 0.5$ occurs when the inferred orientation of the lower pole is toward the general direction of P3-P4. We surmise that the increased magnetic pressure on that part of the stream is able to balance the decreasing ram pressure, resulting in a luminous threading region. Since the P3-P4 vicinity is in the $-y$ half of Figure~\ref{model}, an emission region there would result in an early ingress. In all likelihood, the centroid of that threading region does not approach P4 or its midpoint because the theoretical O$-$C values do not agree with the observed values. However, a centroid closer to P3 would result in a less-early eclipse which would be more consistent with the observations.

As the WD slowly rotates clockwise in Figure~\ref{model}, the corresponding changes in the magnetic pressure along the stream's ballistic trajectory would move the position of the threading region within the binary rest frame, and the eclipses would gradually shift to later phases. Half a beat cycle after the  $\phi_{beat} \sim 0.5$ jump in O$-$C timings, the lower pole would be oriented in the general direction of P2 and the upper pole towards P4. As the upper pole's magnetic pressure increases on the stream in the P3-P4 vicinity, a new threading region would form there, producing the O$-$C jump observed near $\phi_{beat} \sim 0.0$. In short, our model predicts the two distinct O$-$C jumps and explains why they are from late eclipses to earlier eclipses.

Our observations provide circumstantial evidence of the brief, simultaneous presence of two separate emission regions as the system undergoes its O$-$C jump near $\phi_{beat} \sim 0.5$ during one beat cycle in July 2014. On JD 2456842, less than one day before the O$-$C jump, the time of minimum eclipse flux had an O$-$C of $\sim$140 seconds, but on the very next night, there were two distinct minima within the same eclipse. Separated by a prominent increase in brightness, one minimum had an O$-$C of $-80$ seconds, while the other had an O$-$C of 240 seconds, consistent with the presence of discrete emission regions in the $-y$ and $+y$ halves of the plot in Figure~\ref{model}. Moreover, assuming a WD eclipse duration of 700 seconds \citep{mukai} centered upon orbital phase 0.0, the optical eclipse on the first night commenced when the donor occulted the WD, implying a lack of emission in the $-y$ region. However, the egress of that eclipse continued well after the reappearance of the WD, as one would expect if there were considerable emission in the $+y$ area. Indeed, a centroid of emission near the P1 midpoint would account for the observed O$-$C value. On the ensuing night, by contrast, the eclipse began before the disappearance of the WD, and ended almost exactly when the WD reappeared. The implication of these two light curves is that within a 24-hour span between $\phi_{beat} \sim 0.47-0.48$, the locations of the emission regions changed dramatically. Figure~\ref{shift} shows these light curves and indicates in both of them the times of anticipated WD ingress and egress. Further observations are necessary to determine whether this behavior recurs during each beat cycle.

\begin{figure*}

	\begin{subfigure}{
	\includegraphics[width=0.5\textwidth]{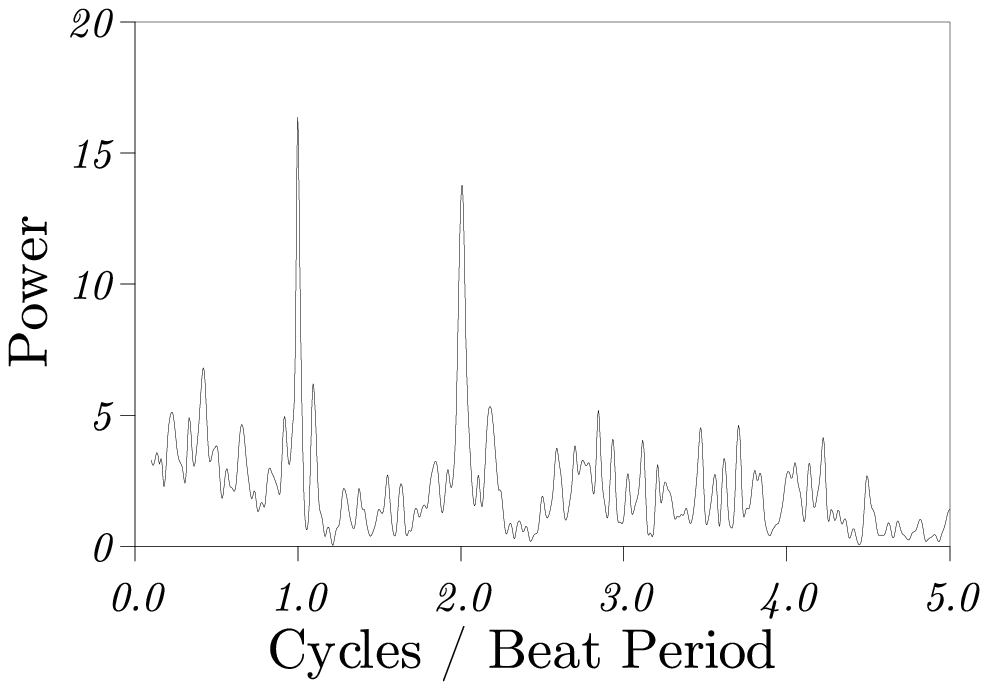}
	\includegraphics[width=0.5\textwidth]{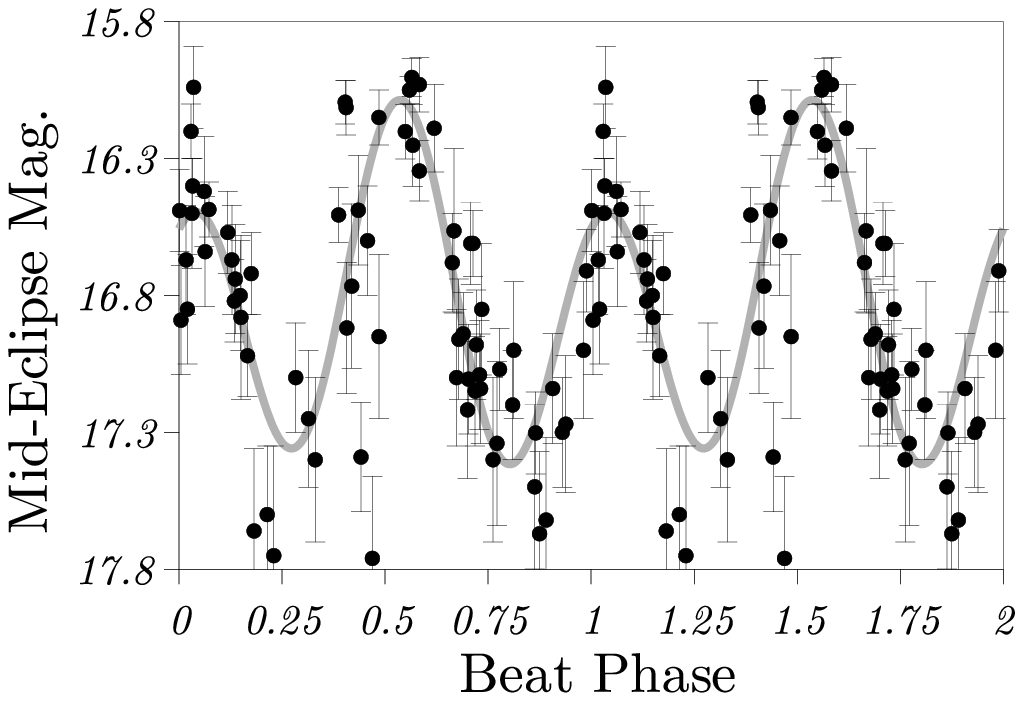}}
	\end{subfigure}
	
\caption{The power spectrum of the residual flux and a phase plot showing the waveform of the signal at the beat period. Spanning 11.8 beat cycles, these plots use only the observations made with the 28-cm Notre Dame telescope. The double-wave sinusoid in the phase plot is meant to assist with visualizing the data and does not represent an actual theoretical model of the system. }
\label{eclipses}
\end{figure*}

\subsubsection{Implications of Findings}

Our hypothesis that the location of the threading radius is variable has ramifications for previous works. In particular, \citet{gs97} and \citet{staubert03} used the timing residuals of the spin minima to track the accretion spot as it traced an ellipse around one of the magnetic poles. One of their assumptions was that the threading radius is constant, but this is inconsistent with the conclusions we infer from our observations and model of the system. A variable threading radius would change the size and shape of the path of the accretion spot \citep{mukai88}---and therefore, of the waveform of the spin minima timings used in those studies to constrain the accretion geometry.

Additionally, the agreement between the model and our observations provides compelling evidence which substantiates previous claims (see Section~\ref{intro}) that the accretion stream in V1432 Aql is able to travel around the WD, as is also observed in the other asynchronous polars. The inefficient threading in asynchronous systems could be indicative of a relatively weak magnetic field or a high mass-transfer rate. For example, \citet{schwarz} found that if the accretion rate in the asynchronous polar BY Cam were 10-20 times higher than normal accretion rates in polars, the stream could punch deeply enough into the WD's magnetosphere to reproduce the observed azimuthal extent of the accretion curtain. Although it is at least conceivable that the asynchronism itself causes the inefficient threading, it is not immediately apparent why this would be so when $P_{sp}$ and $P_{orb}$ are so close to each other.

Regarding the possibility of a high mass-transfer rate, previous works \citep[e.g.,][]{kps88} have proposed that irradiation by a nova can temporarily induce an elevated mass-transfer rate which persists for many decades after the eruption has ended. In line with this theory, \citet{bklyn} proposed that CVs with consistently elevated mass-transfer rates---specifically, nova-like and ER UMa systems---exist fleetingly while the donor star cools after having been extensively irradiated by a nova. If all asynchronous polars are recent novae, as is commonly believed, this theory would predict that the same nova which desynchronizes the system also triggers a sustained, heightened mass-transfer rate as a result of irradiation. The increased ram pressure of the accretion stream would enable it to penetrate deeply into the WD's magnetosphere, thereby offering a plausible explanation as to why all four confirmed asynchronous polars show strong observational evidence of inefficient threading. However, this would not resolve the problem of the short nova-recurrence time in polars \citep[][ discussed in Section~\ref{intro}]{warner02}.

\subsection{Variations in the Residual Eclipse Flux}

\subsubsection{Periodicity} \label{flux-periodicity}

The WD is invisible during eclipse, leaving two possible causes for the variation in residual eclipse flux: the donor star and the accretion stream. The magnetic field lines of the WD can carry captured material above the orbital plane of the system, so depending on projection effects, some of the accretion flow could remain visible throughout the WD's eclipse. Therefore, as the accretion flow threads onto different magnetic field lines throughout the beat period, the resulting variations in the accretion flow's trajectory could cause the residual eclipse flux to vary as a function of $\phi_{beat}$. 

After we calculated the beat cycle count ($C_{beat}$) for each eclipse observation, we generated a power spectrum using the ANOVA method with $C_{beat}$ as the abscissa and the minimum magnitude as the ordinate. For this particular periodogram, we used only the 71 eclipses observed with the 28-cm Notre Dame telescope due to the difficulty of combining unfiltered data obtained with different equipment. The strongest signal in the resulting power spectrum has a frequency of $0.998 \pm0.012$ cycles per beat period. Figure~\ref{eclipses} shows both the periodogram and the corresponding phase plot, with two unequal maxima per beat cycle.

While a double-wave sinusoid provides an excellent overall fit to the residual-flux variations, the observed mid-eclipse magnitude deviated strongly from the double sinusoid near $\phi_{beat} \sim 0.47$ in at least three beat cycles.\footnote{While there are sporadic departures from the double-sinusoid, none is as dramatic as the behavior near $\phi_{beat} \sim 0.47$ or shows evidence of persistence across multiple beat cycles.} Two eclipses observed on consecutive nights in high-cadence photometry with the 80-cm Krizmanich telescope provide the best example of this variation. On JD 2456842, the system plummeted to $V\sim17.8$ during an eclipse ($\phi_{beat} = 0.469$) near the expected time of maximum residual flux. But just 24 hours later, the mid-eclipse magnitude had surged to $V\sim16.2$ ($\phi_{beat} = 0.485$), which was the approximate brightness predicted by the double-sinusoid fit. Furthermore, the eclipse light curve from the second night exhibited intricate structure which had not been present during the previous night's eclipse. These light curves were shown in Figure~\ref{shift}. Comparably deep eclipses near $\phi_{beat}\sim0.47$ were observed during two additional beat cycles (one in 2013 and another in 2014), so there is at least some evidence that the residual flux might be consistently lower near this beat phase. Unfortunately, gaps in our data coverage make it impossible to ascertain whether the mid-eclipse magnitude always fluctuates near $\phi_{beat} \sim 0.47$, so confirmation of this enigmatic variation is necessary.

\subsubsection{Application of Model}

\begin{figure}

	\centerline{\includegraphics[width=0.45\textwidth]{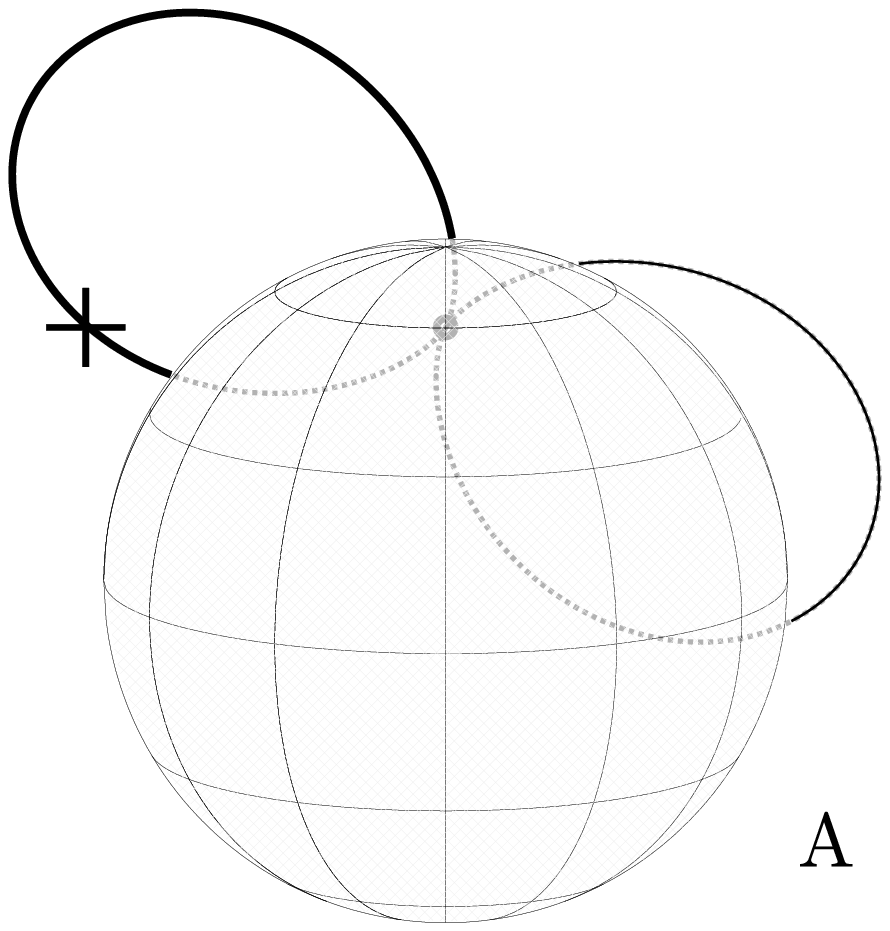}}
	\par
	\centerline{\includegraphics[width=0.45\textwidth]{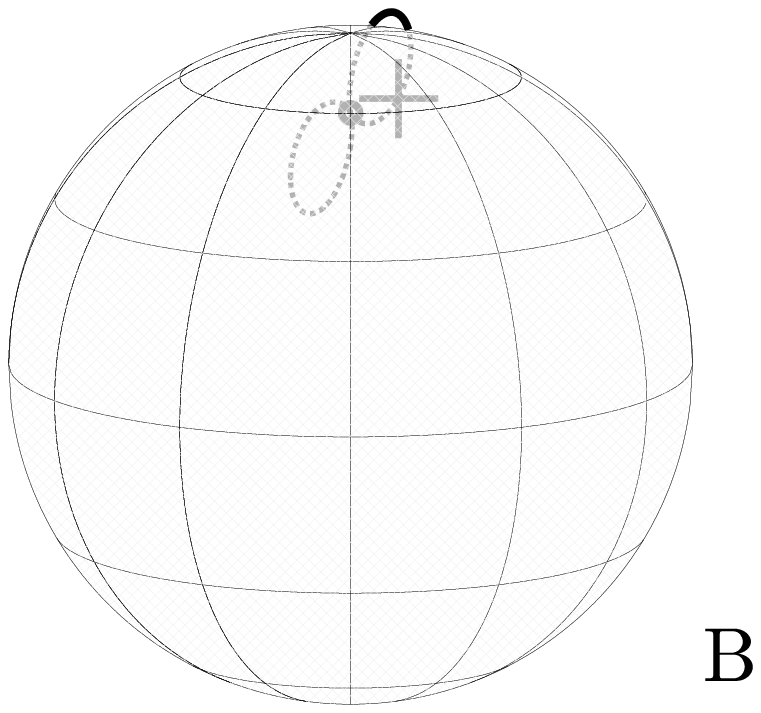}}	

\caption{Two schematic diagrams providing a simplified illustration of our explanation for the residual flux variations at mid-eclipse. In both panels, the captured material travels in both directions along an illustrative magnetic field line. The secondary is the gray sphere eclipsing the WD, and the threading point is shown as a large $+$. The inclination of the magnetic axis with respect to the rotational axis was arbitrarily chosen as 30$^{\circ}$. The portion of the magnetic stream which travels upward and which is visible at mideclipse is highlighted. The threading point in Panel A is near P4, and its threading radius is 3.6 times larger than that of the threading point in Panel B, when the threading point is near the stream's closest approach to the WD.}

\label{flux-diagram}
\end{figure}

We propose that the overall variation in mid-eclipse flux is the signature of an accretion curtain whose vertical extent varies as a function of the threading radius. When the threading region is farther from the WD, the stream can couple onto magnetic field lines which achieve such a high altitude above the orbital plane that the donor star cannot fully eclipse them. By contrast, when the threading region is closer to the WD, the corresponding magnetic field lines are more compact, producing a smaller accretion curtain which the donor occults more fully. The schematic diagram in Figure~\ref{flux-diagram} offers a visualization of this scenario.

While it is conceivable that the residual flux variation is caused by material within the orbital plane, the available evidence disfavors this possibility. In particular, \citet{ss01} saw no diminution in the strength of high-excitation UV emission lines during an eclipse with considerable residual flux at $\phi_{beat} = 0.58$. If these emission lines originated within the orbital plane, they would have faded during the eclipse. Furthermore, if the source of the residual flux were in the orbital plane, the eclipse width would likely correlate with the mid-eclipse magnitude. The eclipses with high levels of residual flux would be long, while the deeper eclipses would be short. We do not see this pattern in our data, and Figure~12 in \citet{boyd} does not show such a correlation, either.

Our model from Section~\ref{description_of_model} predicts that the threading radius will vary by a factor of $\sim3.6$ between P4 and the stream's point of closest approach to the WD. (We reiterate that since these points are meant to be illustrative, this is not necessarily the actual variation in the threading radius.) The upshot is that at P3, threading would take place significantly deeper in the WD's magnetosphere than it would at P4. Moreover, since the predicted threading radius would be largest near an O$-$C jump, this hypothesis predicts that the amount of residual flux would be greatest near those jumps and lowest between them, as is observed in a comparison of Figures~\ref{timing}~and~\ref{eclipses}. In the case of a magnetic stream originating from a threading region between P2-P4, the midpoint of the stream would be visible if it achieves a minimum altitude of $z \sim 0.08a$ above the orbital plane, where $a$ is the binary separation. At P4, this is only one-quarter the predicted threading radius, but at P2 and P3, this is three-quarters of the predicted threading radius. 

This hypothesis also explains why some spectra of V1432 Aql during mid-eclipse show intense emission lines \citep[e.g.][]{watson, ss01}, while others show only weak emission \citep[e.g.][]{patterson}. For each of these previously published spectroscopic observations, we calculated $\phi_{beat}$ and found that the ones showing strong emission lines were obtained when the predicted residual flux was near one of its maxima in Figure~\ref{eclipses}. By contrast, the spectra containing weak emission were obtained when the expected residual flux was approaching one of its minima. If our hypothesis is correct, then the variation in the emission lines is simply the result of the changing visibility of the accretion curtain during eclipse. \citet{watson} suggested a somewhat related scenario to account for the presence of emission lines throughout the eclipse, but they disfavored this possibility largely because of the apparent residual flux at X-ray wavelengths. (As mentioned previously, \citet{mukai} later demonstrated that the residual X-ray flux was contamination from a nearby galaxy.)

An excellent way to test our theory would be to obtain Doppler tomograms near the times of maximum and minimum residual eclipse flux. \citet{schwarz} showed that this technique is capable of revealing the azimuthal extent of the accretion curtain in BY Cam, and it would likely prove to be equally effective with V1432 Aql.

We do not have enough data to consider why the residual flux can vary by as much as $\sim$1.5 mag in one day near the expected time of maximum residual flux. Knowing whether the residual flux is always low near $\phi_{beat} = 0.47$ would be a necessary first step in this analysis.

\subsection{The Dependence of the Spin Modulation on Beat Phase}\label{spin}

As the WD slowly spins with respect to the secondary, the accretion stream will couple to different magnetic field lines, meaning that the spin modulation will gradually change throughout the beat cycle. To explore this variation, we constructed non-overlapping, binned phase plots of the spin modulation in ten equal segments of the beat cycle ({\it e.g.}, between $0.00 < \phi_{beat} < 0.10$). As with the residual-eclipse-flux measurements, we used only the data obtained with the Notre Dame 28-cm telescope in order to avoid errors stemming from the different unfiltered spectral responses of multiple telescope-CCD combinations. In an effort to prevent eclipse observations from contaminating the spin modulation, we excluded all observations obtained between orbital phases 0.94 and 1.06. We then calculated the beat phase for all remaining observations and used only those observations which fell into the desired segment of the beat cycle. We used a bin width of 0.01$P_{sp}$, and we did not calculate bins if they consisted of fewer than five individual observations.

\begin{figure*}
	\centering
  	\begin{tabular}{cc}
    \includegraphics[width=.5\textwidth]{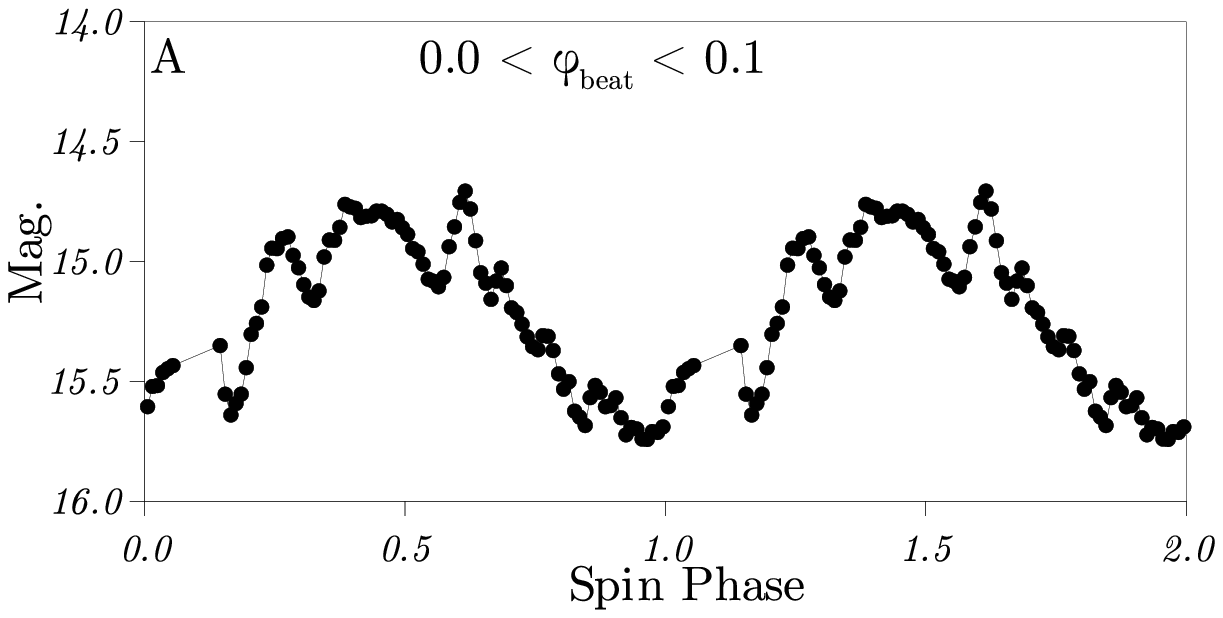} &
    \includegraphics[width=.5\textwidth]{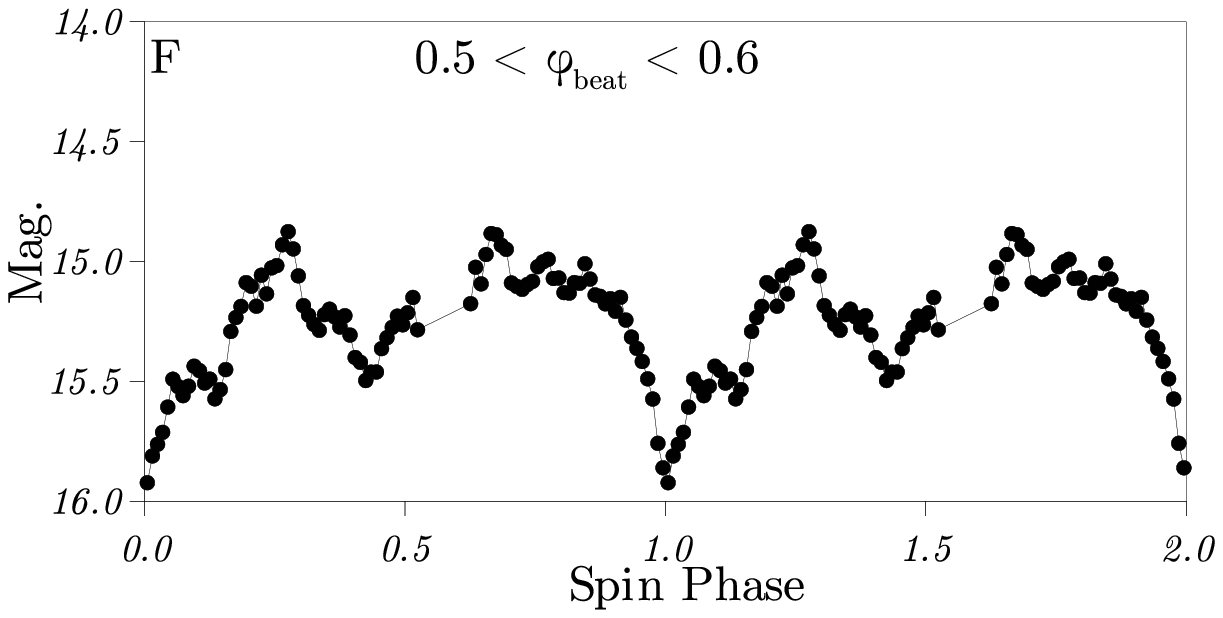} \\
    \includegraphics[width=.5\textwidth]{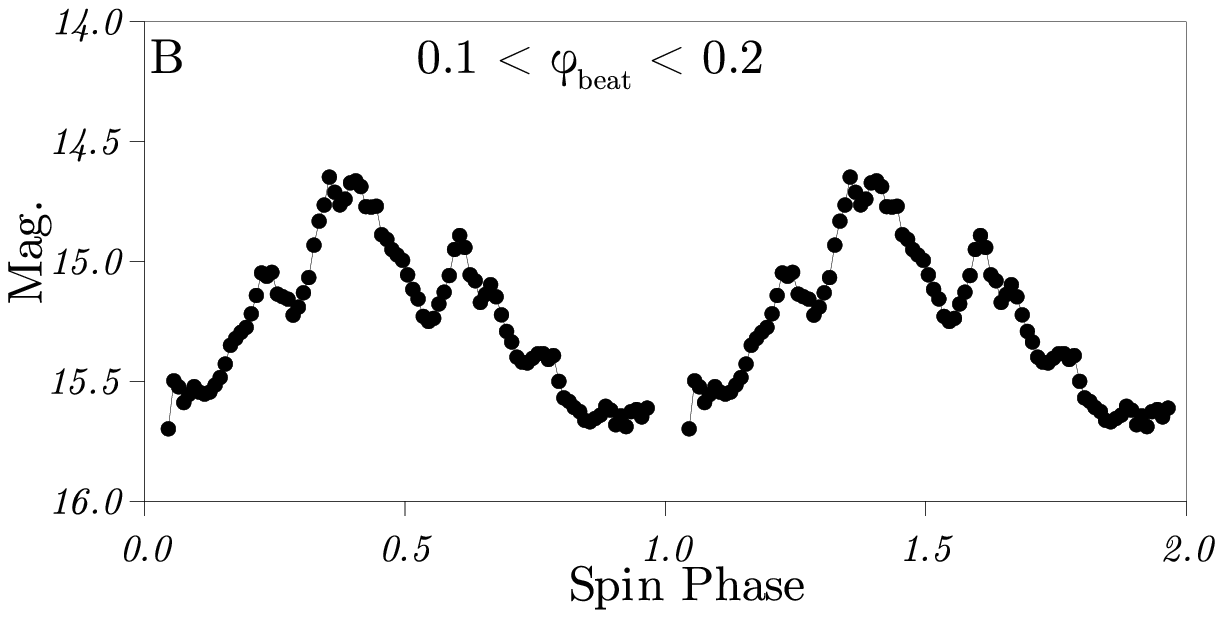} &
    \includegraphics[width=.5\textwidth]{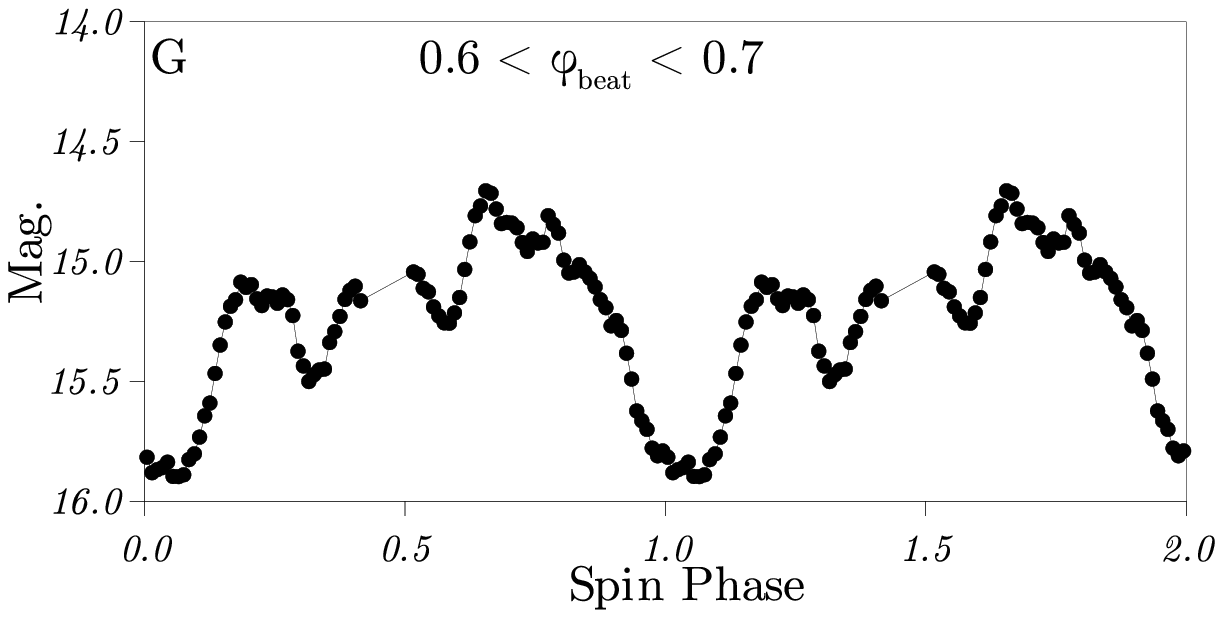}   \\
    \includegraphics[width=.5\textwidth]{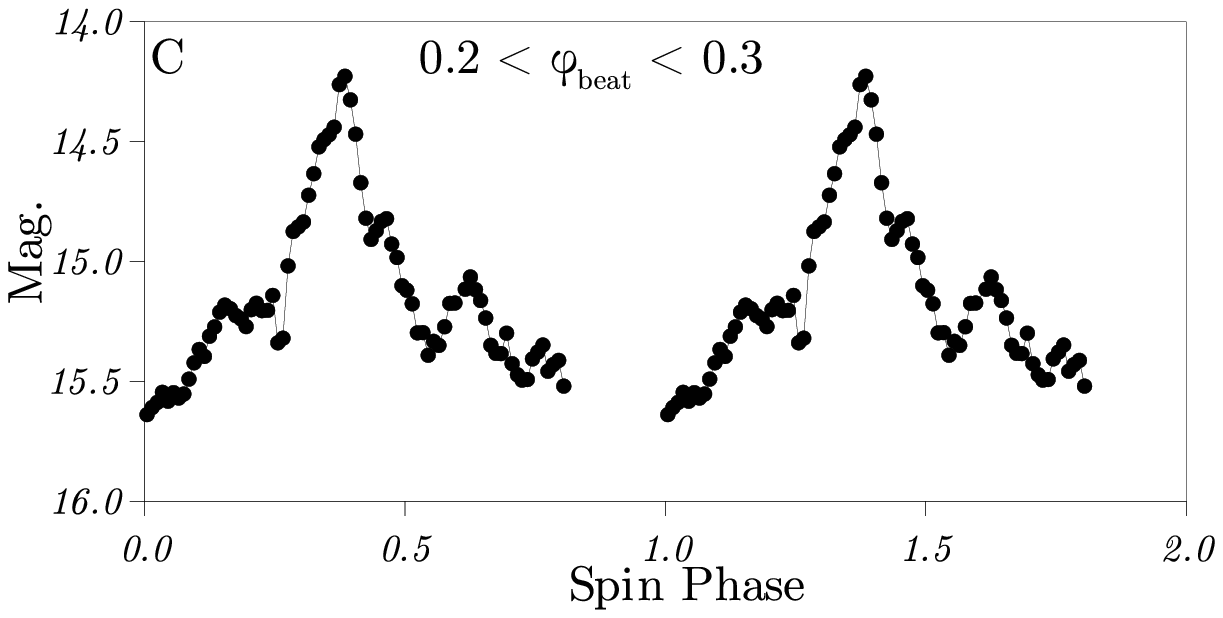} &
    \includegraphics[width=.5\textwidth]{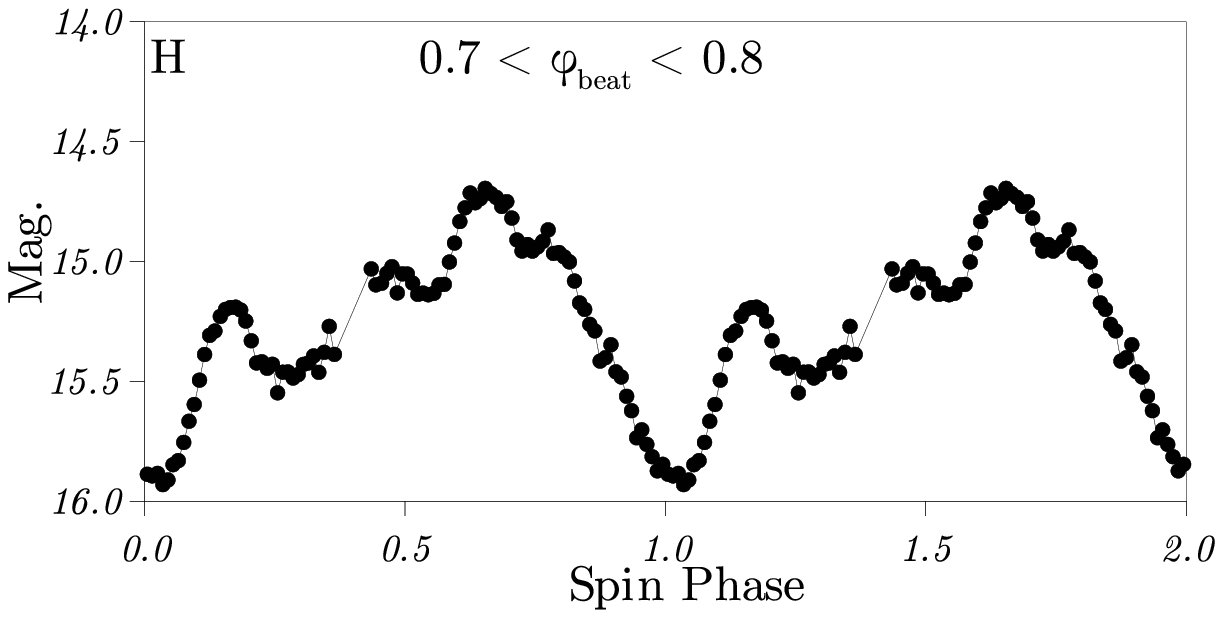} \\
    \includegraphics[width=.5\textwidth]{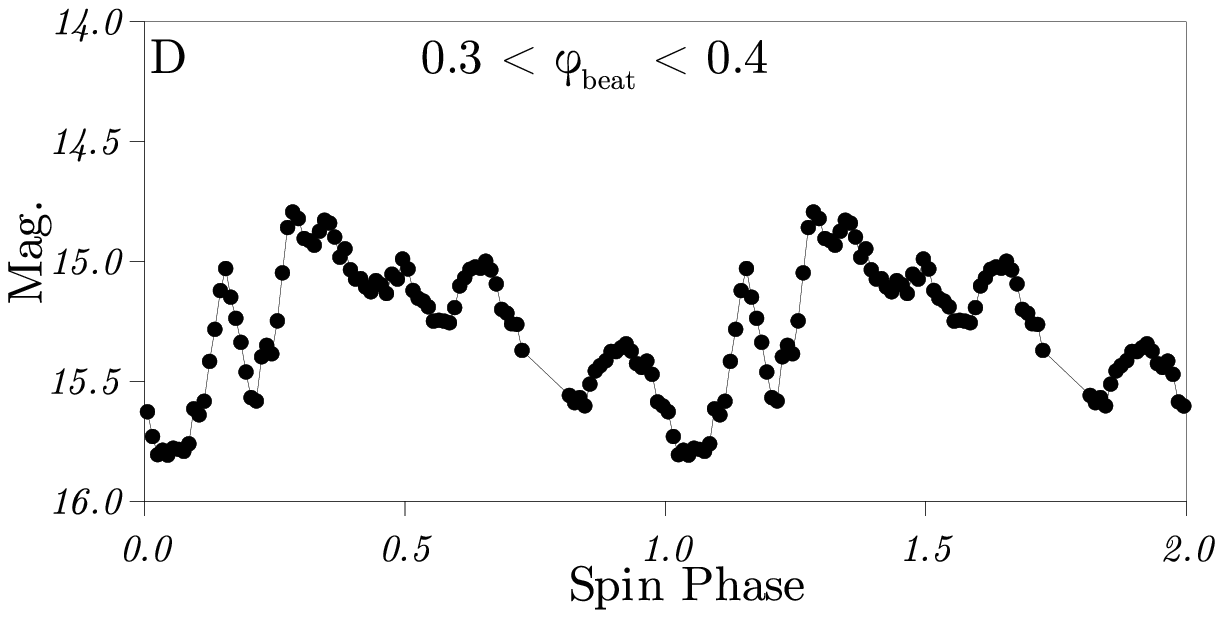} &
    \includegraphics[width=.5\textwidth]{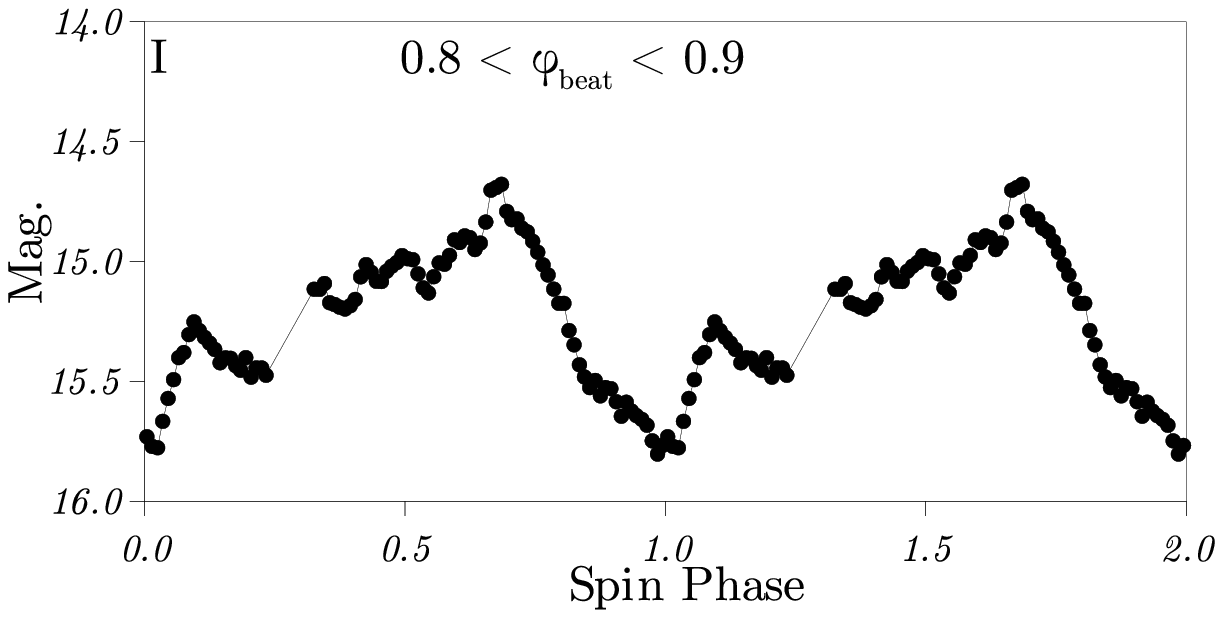}   \\
    \includegraphics[width=.5\textwidth]{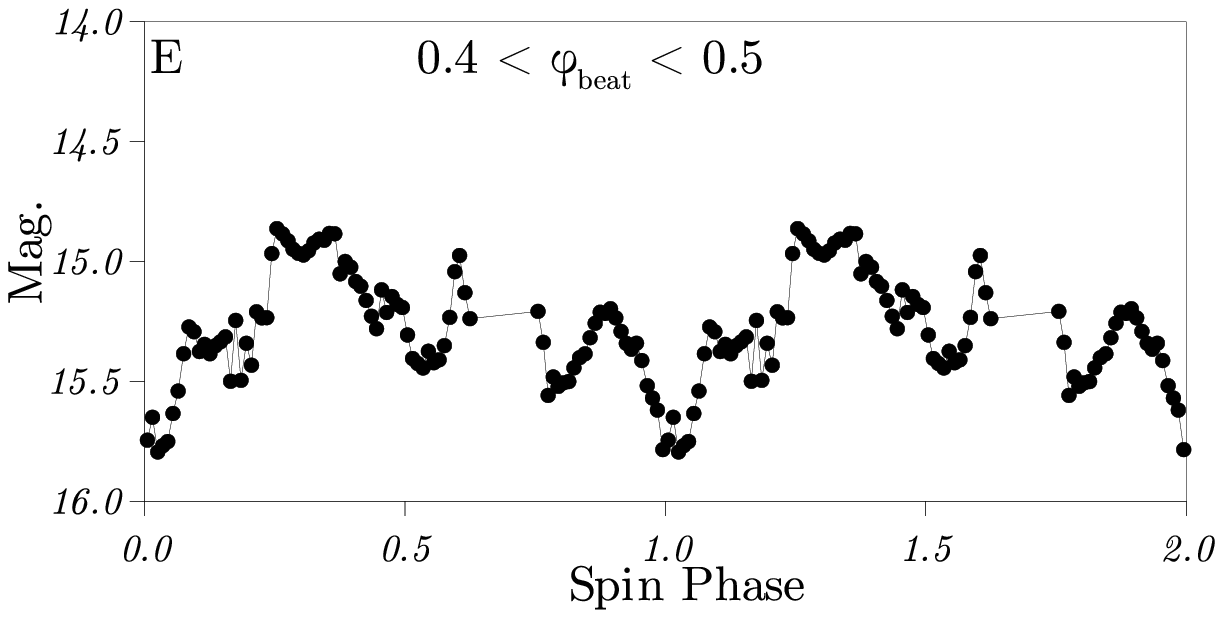} &
    \includegraphics[width=.5\textwidth]{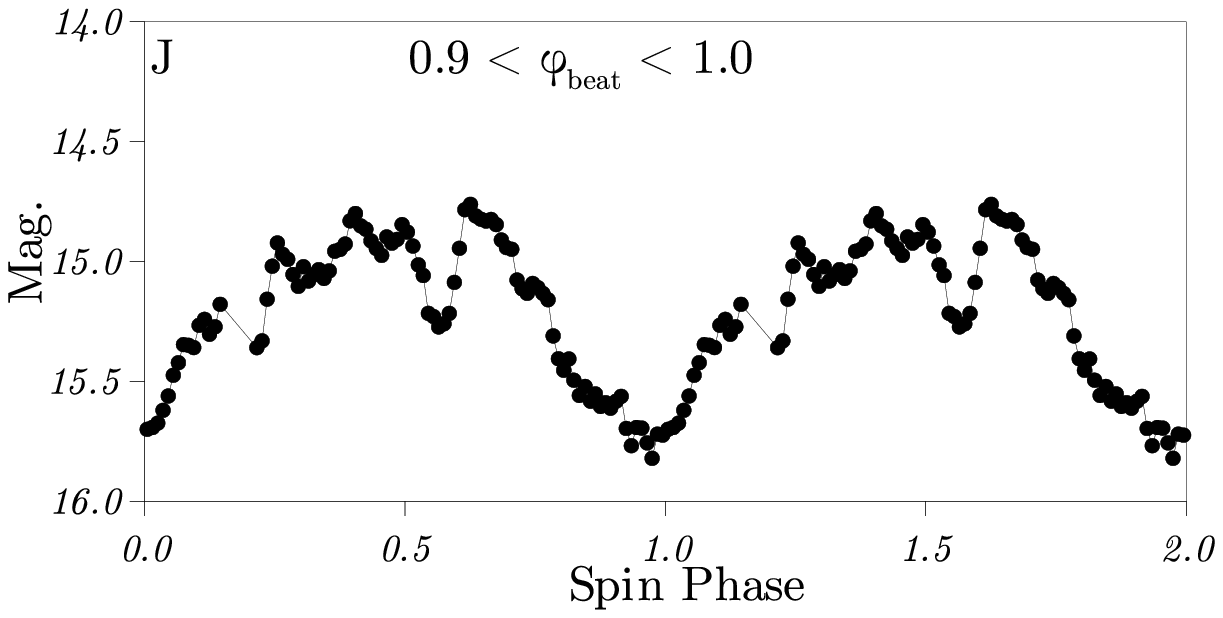} \\
 \end{tabular}
 \caption{Binned phase plots of the spin modulation at different beat phases, with each bin representing 0.01 spin cycles. Gaps in the light curves are due to eclipses. The second spin maximum ($\phi_{sp}\sim0.4$) is strongest in panel C.}
\label{spin-waveform}
\end{figure*}

Figure~\ref{spin-waveform} shows these ten phase plots, and several features are particularly striking. For example, the spin minimum near spin phase 0.0 is highly variable. Conspicuous between $0.5 < \phi_{beat} < 1.0$, it becomes feeble and ill-defined for most of the other half of the beat cycle. Sometimes, the spin minimum is quite smooth and symmetric, as it is between $0.7 < \phi_{beat} < 0.8$, but it is highly asymmetric in other parts of the beat cycle, such as $0.5 < \phi_{beat} < 0.6$. Additionally, there is a striking difference between the phase plots immediately before and after the O$-$C jump near $\phi_{beat} \sim 0.5$, as one would expect if the O$-$C jump marks a drastic change in the accretion geometry.

There is also a stable photometric maximum near spin phase $\sim0.6$ which is visible for most of the beat cycle, though its strength is quite variable. We refer to this feature as the primary spin maximum, but it is not as prominent as the spin minimum. Its behavior is unremarkable.

Interestingly, there is another, much stronger photometric maximum  at $\phi_{sp} \sim 0.4$ which is visible only between $0.0 < \phi_{beat} < 0.5$. Since this feature shares the WD's spin period, we refer to it as the second spin maximum. The second spin maximum can be exceptionally prominent in photometry, attaining a peak brightness of $V \sim 14.1$ in several of our light curves---which is the brightest that we have observed V1432 Aql to be. When visible, the second spin maximum precedes the primary spin maximum by $\sim 0.2$ phase units. It begins to emerge near  $\phi_{beat} \sim 0.0$, and gradually strengthens until it peaks between between $0.2 < \phi_{beat} < 0.3$. It then weakens considerably as $\phi_{beat}$ approaches 0.5, and after the O$-$C jump near $\phi_{beat} \sim 0.5$, the second spin maximum is replaced by a dip in the light curve.

Although the second spin maximum consistently appears between $0.0 < \phi_{beat} < 0.5$, it vanished in a matter of hours on JD 2456842 ($\phi_{beat} \sim 0.47$), only to reappear the next night. On the first night, our observations covered two spin cycles, and while the second spin maximum was obvious in the first cycle, it had disappeared by the second. Just 24 hours later, it was again visible in two successive spin cycles. This unexpected behavior coincides with the approximate beat phase at which we would expect the dominant threading region to shift to the P3-P4 region in our model. Nevertheless, our lack of observations near this beat phase precludes a more rigorous examination of this particular variation.

The second spin maximum is very apparent in some previously published light curves of V1432 Aql from as far back as two decades ago. For example, \citet{watson} presented light curves of V1432 Aql obtained in 1993 which showcase the gradual growth of the second spin maximum (see Panels B-G of their Figure~2). Using our method of determining the beat phase, we extrapolate a beat phase of 0.96 for the light curve shown in their Panel B and a beat phase of 0.12 for the light curve in their Panel G. The increasing strength of the second spin maximum in their light curves agrees with the behavior that we observed at those beat phases (see our Figure~\ref{spin-waveform}). Likewise, Figure~1 in \citet{patterson} shows the second spin maximum at the expected beat phases. These considerations suggest that the second spin maximum is a stable, recurring feature in optical photometry of V1432 Aql.

The overall predictability of the second spin maximum does not answer the more fundamental question of what causes it. One possibility is that it is the result of an elevated accretion rate on one pole for half of the beat cycle. The apparent gap between the two spin maxima, therefore, might simply be the consequence of an absorption dip superimposed on the photometric maximum or a cyclotron beaming effect, splitting the spin maximum into two.

A more interesting scenario is that the second spin maximum could be the optical counterpart to the possible third polecap detected by \citet{rana} in X-ray and polarimetric data. In that study, \citet{rana} detected three distinct maxima in X-ray light curves as well as negative circular polarization at spin phase 0.45, which is the approximate spin phase of the second spin maximum in optical photometry. They also measured positive circular polarization at spin phases 0.1 and 0.7, which correspond with the spin minimum and the primary spin maximum, respectively. Quite fortuitously, the authors obtained their polarimetric observations within several days of the photometric detection of the second spin maximum by \citet{patterson}. Thus, it is reasonable to conclude that the circular polarization feature near spin phase 0.45 is related to the second spin maximum, consistent with a third accreting polecap. 

The conclusions of \citet{rana}, coupled with our identification of a second spin maximum, suggest that V1432 Aql might have at least three accreting polecaps---and therefore, a complex magnetic field. However, the available evidence is inconclusive, and follow-up polarimetry across the beat cycle could clarify the ambiguity concerning the WD's magnetic field structure.

\section{Conclusion}

We have presented the results of a two-year photometric study of V1432 Aql's beat cycle. We have confirmed and analyzed the eclipse O$-$C variations first reported by \citet{gs99}, and we found that the residual mid-eclipse flux is modulated at the system's beat period. We interpret these variations as evidence that the threading region's location within the binary rest frame varies appreciably as a function of beat phase. Doppler tomography of the system at different beat phases could reveal any changes in the azimuthal extent of the accretion curtain, thereby providing a direct observational test of our model of the system.

Our observations provide circumstantial evidence that the mid-eclipse magnitude undergoes high-amplitude variations on a timescale of less than a day near $\phi_{beat} \sim0.47$, deviating strongly from the expected brightness at that beat phase. In the most remarkable example of this variation, the mid-eclipse magnitude varied by $\sim$1.5 mag in two eclipses observed just 24 hours apart. Whereas the first eclipse was deep and smooth, the second eclipse was shallow and W-shaped, with two distinct minima. Similar variations in residual flux were observed in two other beat cycles, providing at least some evidence that this behavior might be recurrent. Still, additional photometric observations are necessary to confirm the $\phi_{beat}\sim0.47$ fluctuations in mid-eclipse magnitude. Amateur astronomers are ideally suited to undertake such an investigation, especially when one considers that our residual-flux analysis utilized a small telescope and commercially available CCD camera. Moreover, observers with larger telescopes could also obtain relatively high-cadence photometry to study whether double-minima eclipses consistently appear near this beat phase.

In addition, we report a second photometric spin maximum which appears for only about half of the beat cycle. This phenomenon might be evidence of a complex magnetic field, but a careful polarimetric study of the beat cycle would be necessary to investigate this possibility in additional detail.

We also offer updated ephemerides of the orbital and spin periods (see Sec.~\ref{ephem}), as well as a Python script which calculates V1432 Aql's beat phase at a given time and which also predicts when the system will reach a user-specified beat phase. An exponential spin ephemeris models the data as well as a polynomial ephemeris and is consistent with an asymptotic approach of the spin period toward the orbital period. According to the exponential ephemeris, the rate of change of the spin period is proportional to the level of asynchronism in the system; consequently, if the exponential ephemeris were to remain valid indefinitely, the resynchronization process in V1432 Aql would take considerably longer than previous estimates.

Finally, while a comprehensive theoretical model of V1432 Aql is beyond the scope of this paper, such an analysis could refine our description of the system and shed additional light on V1432 Aql's unusual threading mechanisms.

\section*{Acknowledgments}

We thank Peter Garnavich and Joe Patterson for their helpful comments, as well as the anonymous referee, whose suggestions greatly improved the paper.

This study made use of observations in the AAVSO International Database, which consists of variable star observations contributed by a worldwide network of observers.

The Sarah L. Krizmanich Telescope was generously donated to the University of Notre Dame in memory of its namesake by the Krizmanich family. This is the first publication to make use of data obtained with this instrument.

DB, MC, and JU participate in the Center for Backyard Astrophysics collaboration, which utilizes a global team of professional and amateur astronomers to study cataclysmic variable stars.

\appendix

\section[]{Determining the Beat Phase} \label{beatphase}

The spin-orbit beat cycle is the key to making sense of V1432 Aql's behavior. To calculate the beat phase ($\phi_{beat}$) of an observation is to determine the relative orientation of the WD's magnetic field at that time. However, since the WD spin period is variable, the beat period ($P_{beat}$) changes, too. For example, \citet{patterson} measured $P_{sp}$ = 12150 seconds, leading to a $P_{beat}\sim$50 days. But by 2013, the spin period had decreased, leading to a beat period of $\sim$62 days. This Appendix outlines the procedure that we employ in our Python script to calculate $\phi_{beat}$ given the time of observation ($T$).

Since $P_{beat}$ is given by \begin{equation}P^{-1}_{beat} = |P^{-1}_{orb}-P^{-1}_{sp}|,\label{beat_period_def}\end{equation} one solution is to determine the average length of the spin period ($\bar{P}_{sp}$) between $T$ and $T_{0}$. The first step in determining $\bar{P}_{sp}$ is to differentiate the cubic spin ephemeris from Section~\ref{ephem} with respect to the spin epoch $E_{sp}$, yielding a formula for the instantaneous spin period. $\bar{P}_{sp}$ is given by \begin{equation}\bar{P}_{sp} = \frac{1}{E_{T}}\int_0^{E_{T}} (P_{0}+\dot{P}E_{sp}+\frac{1}{2}\ddot{P}E^2_{sp}) dE,\end{equation} where $E_{T}$ is the number of spin cycles between $T$ and $T_0$. $E_{T}$, in turn, is found by applying the cubic formula to the spin ephemeris in order to express $E$ as a function of $T$.

Once known, $\bar{P}_{sp}$ may be used in conjunction with $P_{orb}$ in Equation~\ref{beat_period_def} to determine the average length of the beat period ($\bar{P}_{beat}$) between $T$ and $T_{0}$. Thus, the number of beat cycles since $T_0$ is  \begin{equation}C_{beat} = \frac{T-T_{0}}{\bar{P}_{beat}},\end{equation} the decimal portion of which is $\phi_{beat}$. In our beat-phase calculations, we arbitrarily selected $T_{0} = 2449638.3278$ from our cubic spin ephemeris, so at $\phi_{beat} = 0.0$, the spin phase is 0.0 and the orbital phase is 0.86.

\bsp
\label{lastpage}

\end{document}